\documentclass[pre,amsmath,aps,twocolumn,superscriptaddress]{revtex4} 

\usepackage{graphicx,xspace}
\usepackage[dvipsnames]{xcolor}
\usepackage{microtype}
\usepackage{hyperref}
\usepackage{siunitx}
\usepackage{amsmath}
\usepackage[normalem]{ulem}
\usepackage{booktabs}
\hypersetup{colorlinks=true,
    linkcolor=blue,
    urlcolor=blue,
    citecolor=blue}
\usepackage{xr}
\usepackage{bm}
\usepackage{doi}

\usepackage[utf8]{inputenc}
\usepackage{textcomp}

\begin{document}

\title{Molecular-to-polymeric crossover in ion diffusion in glyme-based electrolytes: \\
from vehicular to hopping transport}

\author{Aicha Jani}

\affiliation{Université Paris-Saclay, CNRS, LPS, 91405 Orsay, France}

\author{Simon Gravelle}
\email{simon.gravelle@cnrs.fr}

\affiliation{Université Grenoble Alpes, CNRS, LIPhy, 38000 Grenoble, France}

\author{Pawel Wzietek}
\affiliation{Université Paris-Saclay, CNRS, LPS, 91405 Orsay, France}

\author{Mehdi Zeghal$^{\dagger}$}

\affiliation{Université Paris-Saclay, CNRS, LPS, 91405 Orsay, France}

\author{Patrick Judeinstein}

\affiliation{Université Paris-Saclay, CNRS, LPS, 91405 Orsay, France}
\affiliation{Université Paris-Saclay, CNRS, CEA, LLB, 91191 Gif-sur-Yvette, France}

\begin{abstract}
\noindent Ion transport in glyme-based electrolytes arises from a complex interplay between solvation structure, ion correlations, and polymer chain length. Here, combining pulsed-field gradient nuclear magnetic resonance (PFG-NMR), ionic conductivity measurements, and molecular dynamics (MD) simulations, we investigate the diffusion of monovalent cations (Li$^+$, Na$^+$, Cs$^+$) and TFSI$^-$ anions across a wide molecular-weight range, from monoglyme to long poly(ethylene oxide) (PEO) chains up to 4000~g/mol, corresponding to $n$ up to 88, where $n$ is the number of ethylene oxide repeat units. We identify a crossover region at $n \approx 8$ separating two transport regimes. For short chains, ion motion is consistent with a vehicular mechanism, accompanied by pronounced ion correlations. For longer chains, ion transport decouples from polymer motion and proceeds via rapid coordination exchanges within a slowly relaxing matrix. This transition is accompanied by reduced ion clustering and enhanced anion mobility, leading to increasingly anion-dominated charge transport. Overall, our results provide a molecular picture of ion transport across the molecular-to-polymeric transition and highlight the central role of solvation shell dynamics and polymer relaxation in governing ion dynamics in glyme-based electrolytes.
\end{abstract}

\maketitle

$\dagger$ in memoriam of our colleague Mehdi Zeghal

\section{Introduction}

\noindent Ion transport in liquid and polymer electrolytes remains an open question in soft condensed matter and energy storage physics. In polymer-based electrolytes, ion motion is governed by a complex interplay between segmental polymer dynamics, transient ion pairing, and the formation of ion clusters within a dynamically evolving solvation environment~\cite{Armand1979,Armand1980,gray1991solid,hallinanPolymerElectrolytes2013}. Despite extensive studies, a molecular-level understanding of how these mechanisms evolve as a function of chain length, from small molecular solvents to fully polymeric matrices, remains elusive.

Poly(ethylene oxide) (PEO) and its oligomeric counterparts, glymes (oligo(ethylene glycol) dimethyl ethers), provide a well-defined model system to address this question~\cite{Armand1978, nicholasImprovedSynthesisOxymethylenelinked1988, devauxMechanismIonTransport2012}. These aprotic, highly polar compounds present multiple ether oxygen coordination sites along the backbone, conferring strong cation-solvating ability while maintaining essentially the same repeat-unit chemistry across a broad range of chain lengths, apart from weak end-group effects associated with the terminal methoxy groups. Such end-group effects are expected to become significant primarily for the shortest oligomers. This chemical continuity allows the molecular-to-polymer transition to be explored within a single framework by varying only the degree of polymerization $n$, making glymes widely used model ion-coordinating media for studying the interplay between solvation, polymer dynamics, and ion pairing~\cite{Fenton1973, Armand1978,berthierMICROSCOPICINVESTIGATIONIONIC1983, chataniCrystalStructurePolyethylene1987, nicholasImprovedSynthesisOxymethylenelinked1988, tobishimaGlymebasedNonaqueousElectrolytes2004,devauxMechanismIonTransport2012,dilecceRechargeableLithiumBattery2016,weiHighPerformanceLithiumMetalBattery2020,dilecceGlymebasedElectrolytesSuitable2022}.

\begin{figure}
\centering
\includegraphics[width=\columnwidth]{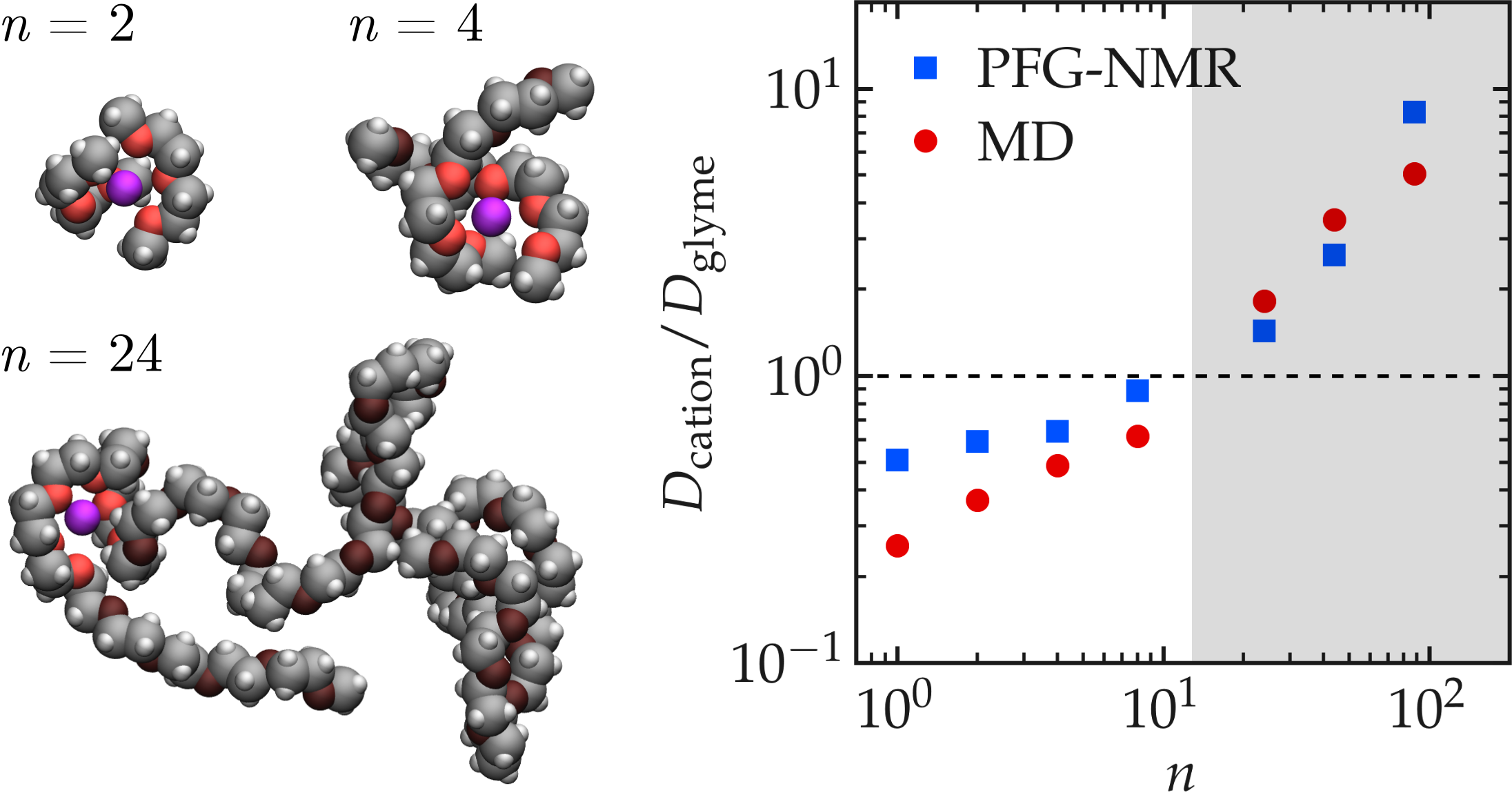}
\caption{Left: Representative MD simulation snapshots of the Cs$^+$ solvation environment for $n = 2$, $n = 4$, and $n = 24$. The Cs$^+$ ion is shown in pink, PEO carbon and hydrogen atoms in gray, and PEO oxygen atoms in red; oxygen atoms outside the first coordination shell (cutoff $r_\mathrm{c} = 4.3$~\AA) are shown in darker red. For $n = 2$ and $4$, the cation is coordinated by oxygen atoms belonging to multiple distinct PEO chains, representative of the short-chain regime. In contrast, for $n = 24$, a single chain wraps around and solvates the cation. Distributions of coordinating oxygen atoms and coordinating chains are provided in Figs.~\ref{supp-fig:environments_Cs}--\ref{supp-fig:environments_Cs_2}. Right: Ratio of the cation (Cs$^+$) self-diffusion coefficient to the glyme self-diffusion coefficient, $D_{\mathrm{cation}}/D_\mathrm{glyme}$, as a function of chain length $n$, from PFG-NMR (blue squares) and MD (red disks). The dashed horizontal line marks $D_{\mathrm{cation}} = D_\mathrm{glyme}$; the shaded region highlights the large-$n$ regime where the cation diffuses faster than the glyme backbone.}
\label{fig:main}
\end{figure}

Yet, despite this apparent simplicity, the relationship between chain length, solvation structure, ion clustering, and macroscopic transport remains poorly understood. Simulations have shown that increasing salt concentration enhances ion clustering and suppresses ionic mobility~\cite{molinariEffectSaltConcentration2018}, while experimental studies have reported seemingly contradictory trends in ion dissociation and conductivity~\cite{berthierMICROSCOPICINVESTIGATIONIONIC1983, hendersonGlymeLithiumBistrifluoromethanesulfonylimideGlymeLithium2005, plewa-marczewskaNMRStudiesEquilibriums2010, mandaiEffectIonicSize2015}. For instance, longer chains are expected to better solvate cations and reduce ion pairing, which should enhance conductivity, yet conductivity decreases monotonically with chain length~\cite{brouilletteStableSolvatesSolution2002, devauxMechanismIonTransport2012}. This apparent paradox reflects a competition between ion dissociation, which improves with chain length, and overall species mobility, which decreases as the chains become longer. A coherent picture linking microscopic ion--polymer interactions to macroscopic transport across the molecular-to-polymeric transition is still missing.

In this work, we address this gap by systematically investigating ion transport in glyme-based electrolytes across a wide molecular-weight range (90--4000~g/mol, $n = 1$ to 88), covering the transition from molecular solvents to polymeric matrices. We examine three alkali salts composed of monovalent cations (Li$^+$, Na$^+$, and Cs$^+$) paired with the bis(trifluoromethanesulfonyl)imide (TFSI$^-$) anion, enabling a direct assessment of cation-identity effects. All measurements are performed at $60~^\circ$C, ensuring that all systems remain in the liquid state. By combining pulsed-field gradient NMR (PFG-NMR) and ionic conductivity measurements with molecular dynamics (MD) simulations, we identify a crossover region at $n \approx 8$ separating two distinct transport regimes (Fig.~\ref{fig:main}), and provide a molecular interpretation in terms of solvation structure, ion clustering, and the competition between ion-polymer residence times and diffusive timescales.

\section{Experimental Details}

\subsection{Materials}
\label{sec:materials}

\noindent Glymes are oligo(ethylene glycol) dimethyl ethers with the general chemical formula $R_1\mathrm{O}(\mathrm{CH_2CH_2O})_n R_2$~\cite{dilecceGlymebasedElectrolytesSuitable2022}. In the present work, $R_1 = R_2 = \mathrm{CH_3}$ and $n = 1$, 2, 4, 8, 24, 44, and 88. Monoglyme (MG, $n=1$), diglyme (DG, $n=2$), tetraglyme (4G, $n=4$), and poly(ethylene glycol) dimethyl ether (P400, $n = 8$) were obtained from Sigma-Aldrich (99~\% purity). Higher-molecular-weight glymes P1100 ($n=24$), P2000 ($n=44$), and P4000 ($n=88$) were purchased from Polymer Source, with polydispersity indices of 1.09, 1.08, and 1.03, respectively (Tab.~\ref{supp-tab:melting-glym}). The possible effect of polydispersity on diffusion is discussed in Sec.~\ref{sec:discussion}.

Electrolyte solutions were prepared at a fixed concentration of 0.8~M using lithium, sodium, or cesium bis(trifluoromethanesulfonyl)imide (TFSI$^-$) salts, with TFSI serving as the common anion (LiTFSI, NaTFSI, and CsTFSI; 99.5~\% purity, Solvionic). This corresponds to an oxygen-to-cation ratio almost equal to 25, defined as the number of glyme ether oxygen atoms per cation. All salts and chemicals were purchased as anhydrous grade materials ($\mathrm{H_2O} < 0.1$~ppm) and handled exclusively in an argon-filled glovebox ($\mathrm{O_2} < 0.1$~ppm, $\mathrm{H_2O} < 0.1$~ppm).

\subsection{PFG-NMR experiments}

\noindent Self-diffusion coefficients were measured by pulsed-field gradient NMR (PFG-NMR) at 9.4~T using a Tecmag system equipped with a 5~mm Doty Scientific diffusion probe capable of generating pulsed field gradients up to 12~T/m. The corresponding resonance frequencies are 400.13, 376.5, 155.5, 105.8, and 52.5~MHz for $^1$H, $^{19}$F, $^7$Li, $^{23}$Na, and $^{133}$Cs, respectively.

Self-diffusion coefficients were measured using the stimulated echo (STE) sequence. The number of scans was adjusted for each system to ensure a sufficient signal-to-noise ratio, with repetition delays corresponding to $5T_1$ (i.e., 15~s for $^1$H and $^{19}$F, 10~s for $^7$Li, 2~s for $^{23}$Na, and 45~s for $^{133}$Cs). The gradient pulse duration $\delta$ was set to 1 or 2~ms, and the gradient strength was adjusted to achieve signal attenuation over approximately two orders of magnitude across the series of experiments. The diffusion time $\Delta$ was varied between 60 and 240~ms depending on the nucleus and solvent. Two values of $\Delta$ were used in selected cases to confirm the absence of convective artifacts. For sodium in monoglyme and diglyme, $\Delta$ was limited to 20~ms and
10~ms, respectively, due to rapid relaxation.

Samples were prepared in a glove box and flame-sealed. NMR tubes of reduced diameter were used to suppress convection, particularly for low-viscosity monoglyme near its boiling point: 2~mm outer diameter tubes for $n \geq 8$, and coaxial double-wall 1-in-2~mm geometry for $n < 8$. Experiments were conducted at $(60 \pm 2)~^\circ$C, and the temperature was calibrated using the self-diffusion coefficient of water~\cite{holzTemperaturedependentSelfdiffusionCoefficients2000} in a reference tube of similar geometry.

The self-diffusion coefficient $D$ was determined from the attenuation of the echo signal $E$ according to \cite{stejskalSpinDiffusionMeasurements1965,priceSelfDiffusionSupercooledWater1999,pricePulsedfieldGradientNuclear1997}:
\begin{equation}
\nonumber
E = \exp\Big[-\gamma^2 g^2 \delta^2 (\Delta - \delta/3) D\Big],
\end{equation}
where $\gamma$ is the gyromagnetic ratio of the nucleus, $g$ is the applied
gradient strength, $\delta$ is the gradient pulse duration, and $\Delta$ is
the diffusion time. Measurements were performed on $^1$H for the solvent,
$^7$Li, $^{23}$Na, or $^{133}$Cs for the cations, and $^{19}$F for the
TFSI$^-$ anion.

Here, 16 equally spaced gradient steps were used for each experiment. Data
acquisition was performed using Tecmag TNMR software, and data processing
was performed using in-house routines developed in Igor Pro.

\subsection{AC impedance spectroscopy experiments}

\noindent AC impedance spectroscopy measurements were carried out using a VMP3 broadband impedance analyzer (BioLogic), over a frequency range from $8.3 \times 10^6$ to $0.83$~Hz. A high temperature conductivity cell (HTCC, BioLogic) was employed, suitable for electrolytes with conductivities ranging from $\SI{2}{\micro\siemens\per\centi\meter}$ to $0.2~\mathrm{S\,cm^{-1}}$. The cell consists of two parallel platinum electrodes and has a nominal cell constant $K = 1.0~\mathrm{cm^{-1}}$, calibrated using standard aqueous conductivity reference solutions.

Electrolyte solutions were prepared and transferred into the conductivity cell in a helium-filled glovebox, with care taken to ensure complete immersion of the electrodes. The sealed cell was then placed in a thermostated oven at $60~^\circ \mathrm{C}$ and allowed to equilibrate for at least 1~h prior to measurement. Temperature stability was monitored using a thermocouple positioned in close proximity to the sample.

The ionic conductivity $\sigma$ was calculated according to
\begin{equation}
\label{eq:impedance}
\sigma = \frac{K}{R},
\end{equation}
where $K$ is the cell constant and $R$ is the bulk resistance obtained from the high-frequency intercept of the Nyquist plot with the real axis (Fig.~\ref{supp-fig:nyquist}). 
The cell constant was calibrated using a 0.1~M KCl standard solution, with conductivities of $11.67~\mathrm{mS\,cm^{-1}}$ at $20~^\circ\mathrm{C}$ and $12.88~\mathrm{mS\,cm^{-1}}$ at $25^\circ\mathrm{C}$. The estimated uncertainty in the conductivity measurements is $\pm 10\,\%$. All impedance measurements were performed at the Institut de Chimie Moléculaire et des Matériaux (ICMMO) laboratory in Orsay, France.

\section{Numerical Details}

\subsection{Molecular dynamics simulations}

\noindent Molecular dynamics (MD) simulations were performed using the GROMACS simulation package~\cite{abrahamGROMACSHighPerformance2015}. Systems consisted of PEO chains of varying degree of polymerization ($n = 1$ to 88), with and without salts (Fig.~\ref{supp-fig:systems}). The number of polymer chains was adjusted such that the total number of atoms in each simulation box was approximately $6 \times 10^3$, ensuring comparable system sizes across all systems (Tab.~\ref{supp-tab:system_composition}). For electrolyte systems cesium bis(trifluoromethanesulfonyl)imide (CsTFSI) salt was added to match the experimental oxygen-to-cation ratio of $\approx 25$.

All simulations employed a force field derived from the CHARMM family, with bonded and non-bonded interactions described by standard Lennard-Jones and Coulombic potentials (Tab.~\ref{supp-tab:ff_parameters}). Polarizable force fields are known to improve the description of glyme-based electrolytes~\cite{liyana-arachchiPolarizableMolecularDynamics2018}, but their computational cost prohibits the microsecond-scale simulations required to reach the diffusive regime for the longest chains considered here. We therefore use a non-polarizable force field, which is a more justified approximation for Cs$^+$ than for small, charge-dense ions such as Li$^+$, since polarization effects scale with charge density. Long-range electrostatic interactions were treated using the particle-mesh Ewald (PME) method~\cite{essmannSmoothParticleMesh1995}, with a real-space cutoff of 1.4~nm and a Fourier grid spacing of 0.1~nm. All scripts for generating molecular topologies, force-field parameters, and GROMACS input files are available on GitHub and archived on Zenodo~\cite{gravelle2026glymesaltmd}.

After energy minimization, systems were equilibrated using successive runs in the NVT and NPT ensembles, at a temperature of 333~K (or $60~^\circ$C) and a pressure of 1~bar similar to those used in the different experiments. Temperature was controlled using a stochastic velocity-rescaling thermostat~\cite{bussiCanonicalSamplingVelocity2007} with a coupling time constant of 0.5~ps, while pressure was maintained using a stochastic cell-rescaling barostat~\cite{bernettiPressureControlUsing2020} with a time constant of 1.0~ps. The equations of motion were integrated using the leap-frog algorithm with a time step of 1~fs. Bond lengths involving hydrogen atoms were constrained using the LINCS algorithm. A Verlet cutoff scheme was used for neighbor searching, with a cutoff radius of 1.4~nm for both van der Waals and Coulomb interactions. Periodic boundary conditions were applied in all directions.

Equilibration was assessed by monitoring the average radius of gyration of the glymes, $R_\text{g}$, as a function of time, and production runs were initiated only after $R_\text{g}$ reached a stable plateau. For the longest chains ($n = 88$), equilibration times up to 500~ns were required. This timescale is comparable to the characteristic diffusive time $\tau \sim R_\text{g}^2/D \approx 400$~ns, estimated using $D \approx 10^{-11}~\mathrm{m^2\,s^{-1}}$ and $R_\text{g} \approx 2$~nm. Production runs were performed in the NPT ensemble, with trajectories extended up to several $\SI{}{\micro\second}$ for the most viscous systems. Configurations were saved every 10~ps for analysis, and center-of-mass motion was removed at each step to prevent system drift.

All analyses, including MSD, clustering, and correlations, were performed from production trajectories using in-house scripts and standard GROMACS tools. Self-diffusion coefficients were extracted from the long-time behavior of the MSD using the Einstein relation (Fig.~\ref{supp-fig:msd}). To ensure a robust identification of the diffusive regime, a sliding-window procedure was applied to the MSD curves in log-log representation. Within each window, the MSD was fitted to a power-law form, and only time intervals exhibiting quasi-linear behavior (slope close to unity) were retained. Among all valid windows, the diffusion coefficient was determined from the optimal time interval selected using a weighting procedure that favors both extended time ranges and slopes closest to unity.

\section{Results}

\subsection{Self diffusion coefficients from PFG-NMR}

\noindent PFG-NMR measurements of the self-diffusion coefficients of glymes, $D_\mathrm{glyme}$, as a function of the degree of polymerization $n$, reveal a systematic decrease with increasing chain length, following a power-law scaling $D_\mathrm{glyme} \propto n^\alpha$ (Fig.~\ref{supp-fig:diffusion-glymes}~A). For pure glymes, the scaling exponent is $\alpha = -1.51$, while in the presence of salt it is $\alpha = -1.55$, indicating that salt addition does not significantly alter the scaling behavior. The diffusion coefficient decreases by approximately a factor of two upon salt addition (Tab.~\ref{supp-tab:label-D-glym}). This reduction is consistent with ion-glyme coordination promoting the formation of transient solvation complexes involving one or more glyme molecules ($n = 2$ and $4$ in Fig.~\ref{fig:main}), which increase the effective hydrodynamic volume of diffusing species. For short glymes, this mechanism likely dominates, whereas for longer chains, hindered segmental mobility becomes increasingly relevant. The identity of the cation (Li$^+$ , Na$^+$ , Cs$^+$) has a negligible effect on glyme diffusion.

\begin{figure}
\centering
\includegraphics[width=\columnwidth]{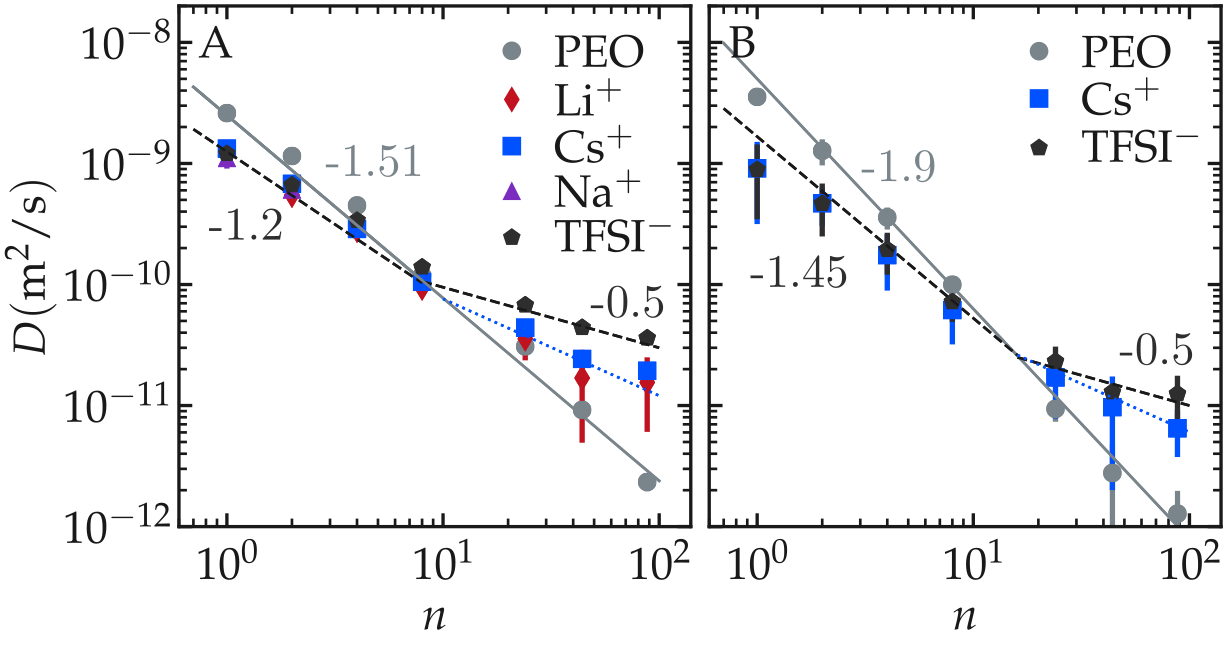}
\caption{A)~Self-diffusion coefficients from PFG-NMR for glyme (gray disks), Cs$^+$ (blue squares),  Na$^+$ (magenta triangles), Li$^+$ (red diamonds), and TFSI$^-$ (black pentagons) as a function of the degree of polymerization, $n$. Guiding lines indicate trends with slopes of -1.51 (solid line), as well as multiple scaling regimes: a slope of $-1.2$ (dashed line) at low $n$, transitioning to slopes of $-0.5$ (dashed line) and $-0.8$ (dotted line) at larger $n$.
All measurements were performed at concentration $C = 0.8$~M and temperature $T = 60~^\circ\mathrm{C}$ (Tab.~\ref{supp-tab:label-D-glym}).
B)~Self-diffusion coefficients from MD for glyme (gray disks), Cs$^+$ (blue squares), and TFSI$^-$ (black stars). Guiding lines indicate trends with slopes of -1.9 (solid line), as well as multiple scaling regimes: a slope of $-1.45$ (dashed line) at low $n$, transitioning to slopes of $-0.5$ (dashed line) and $-0.8$ (dotted line) at larger $n$.}
\label{fig:diffusion}
\end{figure}

The diffusion coefficients of ionic species exhibit a clear change in scaling as a function of the degree of polymerization $n$. For $n \lesssim 8$, cations and anions display similar scaling trends and comparable mobilities. For $n \gtrsim 8$, differences in scaling behavior emerge between anions and cations (Fig.~\ref{fig:diffusion}~A, Fig.~\ref{supp-fig:diffusion_coeff}~A). In particular, TFSI$^-$ becomes increasingly mobile relative to Cs$^+$ and Li$^+$ at large $n$. In addition, a crossover in relative mobility between ions and glymes is observed: ions diffuse more slowly than glymes for short chains, while for long chains, TFSI$^-$ is the most mobile species, followed by cations and glymes. The diffusion coefficients of Li$^+$, Na$^+$, and Cs$^+$ are nearly overlapping, indicating similar dynamics across the different salts (Fig.~\ref{fig:diffusion}~A), although the measurements for Na$^+$ are limited to the shorter glymes ($n \le 2$)~\cite{carboneCharacteristicsGlymeElectrolytes2017, moralesIonTransportAssociation2019} because of the short nuclear relaxation times induced by strong quadrupolar effects, with $T_1$ and $T_2$ relaxation times below 1~ms~\cite{dagostino23NaNMR12021}.

\subsection{Self-diffusion coefficients from MD simulations}

\noindent Molecular dynamics simulations, performed for CsTFSI, reproduce qualitatively the experimental trends (Fig.~\ref{fig:diffusion}, Tab.~\ref{supp-tab:label-D-glym}, Fig.~\ref{supp-fig:diffusion-glymes}-\ref{supp-fig:diffusion_coeff}), showing a similar dependence of self-diffusion coefficients on chain length $n$ and the same crossover behavior at $n \approx 8$. As in experiments, ions exhibit a transition in relative mobility: for short chains ($n \lesssim 8$), ions diffuse more slowly than glymes, whereas for longer chains ($n \gtrsim 8$), TFSI$^-$ becomes the most mobile species, followed by Cs$^+$, with glyme exhibiting the lowest mobility. There are however quantitative differences; MD yields slightly steeper scaling exponents for pure glymes ($\alpha = -1.74$) and glymes with salt ($\alpha = -1.85$), and absolute diffusion coefficients from MD are systematically lower than experimental values (Fig.~\ref{fig:diffusion}, Figs.~\ref{supp-fig:diffusion-glymes}-\ref{supp-fig:diffusion_coeff}).

\subsection{Polymer conformations}

\noindent The radius of gyration of glymes, $R_\mathrm{g}$, was computed from the MD simulations (Fig.~\ref{supp-fig:gyration}). For long chains, $R_\mathrm{g}$ follows the expected scaling $R_\mathrm{g} \propto \sqrt{n}$ (Fig.~\ref{supp-fig:gyration2}). In this limit, the polymer chains behave as flexible random coils, and their conformations are well described by ideal chain statistics~\cite{rubinsteinPolymerPhysics2023}. Deviations are observed at small $n$, which we attribute to finite-size effects induced by the local stiffness of the glymes. For the longest chains considered here, the addition of salt induces a modest reduction in $R_\mathrm{g}$, indicating slight chain compaction, consistent with coordination effects arising from cation solvation, whereby a single cation coordinates multiple ether oxygens along the same chain, effectively favoring more compact conformations ($n = 24$ in Fig.~\ref{fig:main}).

\subsection{Cation environment}

\noindent The local solvation environment of Cs$^+$ ions was characterized by measuring the average number of neighboring oxygen atoms within a cutoff distance $r_\text{c}$ of each cation, where $r_\text{c}$ were defined from the first minimum of the respective radial distribution functions, yielding $r_\text{c} = 4.3$~\AA{} for PEO oxygen atoms and $r_\text{c} = 3.9$~\AA{} for TFSI$^-$ oxygen atoms. For the shortest polymer chains, the coordination environment is shared almost equally between polymer and anion oxygen atoms, with average coordination numbers of approximately $n_\mathrm{O-PEO} \simeq 4$ and $n_\mathrm{O-TFSI} \simeq 3.5$ (Fig.~\ref{supp-fig:environments_Cs}~A). As the chain length increases, a clear reorganization of the solvation shell is observed: $n_\mathrm{O-PEO}$ rises above 6 for $n \ge 8$ while $n_\mathrm{O-TFSI}$ decreases below 2, indicating a progressive suppression of anion participation in the first solvation shell. A decrease in the number of neighboring Cs$^+$ ions with increasing $n$ is also observed (Fig.~\ref{supp-fig:environments_Cs}~B).

The role of polymer connectivity in the solvation structure was further quantified by analyzing the number of distinct PEO chains coordinating each cation (Fig.~\ref{supp-fig:environments_Cs_2}). For short chains, Cs$^+$ ions are frequently coordinated by multiple (typically 2 or 3) polymer chains. In contrast, for long chains with $n > 8$, the distribution of coordinating chains becomes peaked at one, indicating that the majority of cations are solvated by a single PEO chain, as illustrated in Fig.~\ref{fig:main}.

\subsection{Timescales of ion coordination and polymer segmental dynamics}

\begin{figure}
\centering
\includegraphics[width=\columnwidth]{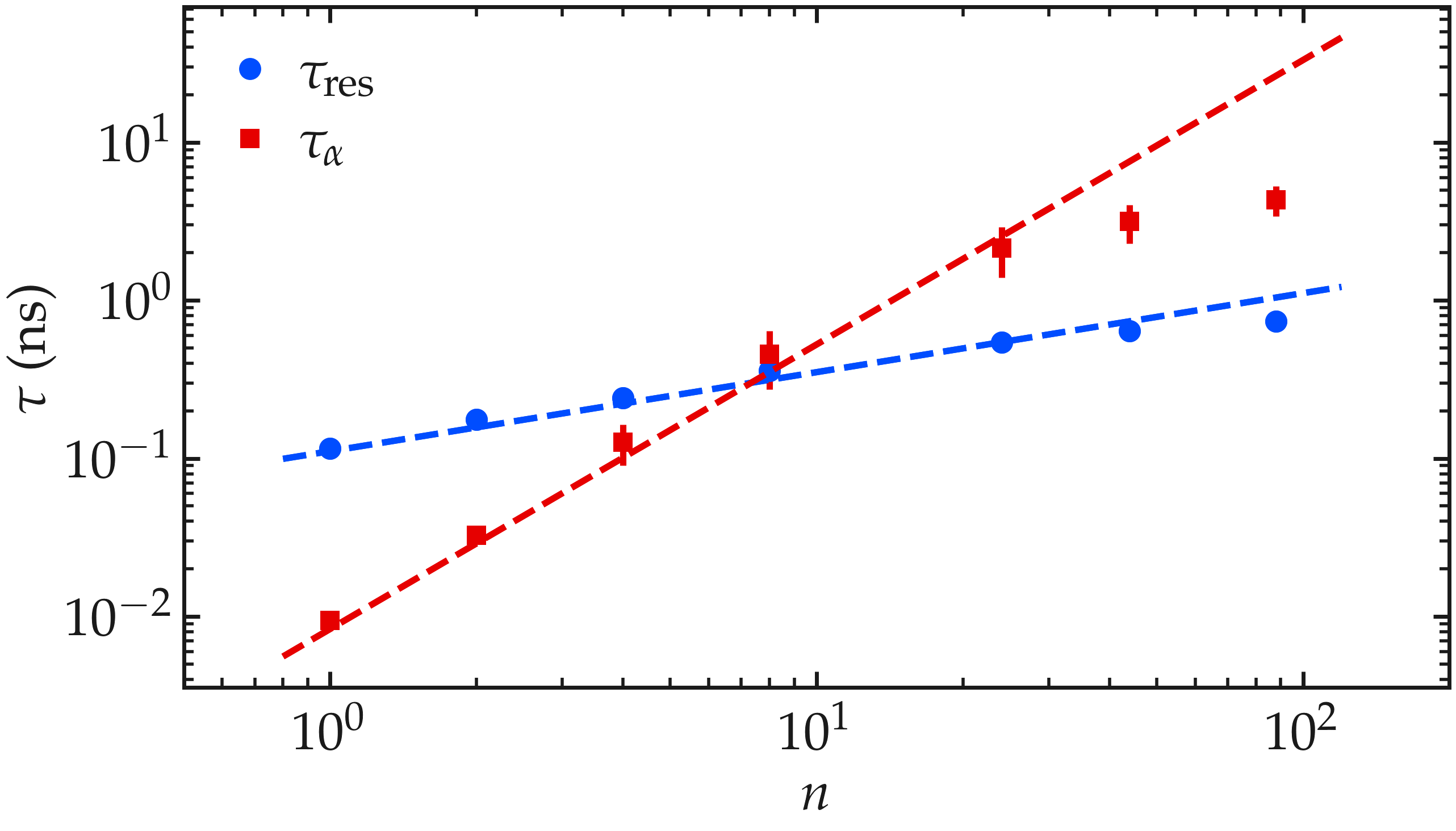}
\caption{Characteristic times as a function of polymer chain length $n$: average residence time of Cs$^+$ ions on a PEO chain, $\tau_\text{res}$ (blue disks), 
and orientational relaxation time of the PEO chains, $\tau_\alpha$ (red squares). Dashed lines are power-law guides $\tau \propto n^{0.5}$ (blue) and $\tau \propto n^{1.8}$ (red).}
\label{fig:times}
\end{figure}

\noindent The average cation residence time on PEO chains, $\tau_\mathrm{res}$, was extracted from MD simulations by tracking whether a cation is coordinated to a given PEO chain --- defined as having at least one chain oxygen within $r_\text{c} = 4.3$~\AA{} --- and analyzing the resulting adsorption--desorption events at the chain level (details are given in the SI). We find that $\tau_\mathrm{res}$ increases monotonically from $\approx 0.1$~ns for $n = 1$ to $\approx 0.7$~ns for $n = 88$ (Fig.~\ref{fig:times}). This moderate increase contrasts with the orientational relaxation time $\tau_\alpha$, extracted from the decay of local PEO bond-vector autocorrelation functions [Eq.~\eqref{supp-eq:bond-vector} from the SI], which probes segmental reorientation dynamics of the polymer backbone and increases by more than two orders of magnitude over the same range of $n$. The two timescales follow markedly different power-law dependencies: $\tau_\mathrm{res} \propto n^{0.5}$ and $\tau_\alpha \propto n^{1.8}$ (for $n \lesssim 24$, Fig.~\ref{fig:times}), indicating a rapid slow down of the polymer dynamics relative to ion exchange kinetics as chain length increases.

Comparing $\tau_\mathrm{res}$ and $\tau_\alpha$ reveals two distinct dynamical regimes (Fig.~\ref{fig:times}). For $n \lesssim 8$, $\tau_\mathrm{res} \gg \tau_\alpha$: polymer segments rearrange rapidly compared to ion exchange events, so ions remain coordinated to their PEO solvation shell while the polymer segments rearrange around them, consequently each ion migrates as part of a persistent solvation complex. For $n \gtrsim 8$, $\tau_\mathrm{res} \ll \tau_\alpha$: ions exchange between PEO chains much faster than the polymer backbone relaxes, enabling transport through successive hopping events along a slowly rearranging matrix. This crossover from an ion-exchange-limited to a polymer-dynamics-limited regime coincides with the change in diffusion scaling observed in Fig.~\ref{fig:diffusion}.

The crossover identified from $\tau_\mathrm{res}$ and $\tau_\alpha$ is further corroborated by an independent estimate of the characteristic time for ionic diffusion over a cation diameter, $\tau_D = \sigma_\mathrm{Cs}^2/(6D)$, where $\sigma_\mathrm{Cs} \approx 0.5$~nm and $D$ is the Cs$^+$ self-diffusion coefficient. Notably, $\tau_D$ and $\tau_\alpha$ follow a similar dependence on $n$ (Fig.~\ref{supp-fig:timescomparison}), indicating that ionic diffusion and polymer segmental relaxation slow down together as chain length increases, which is consistent with a picture where the overall mobility of both ions and polymer segments is controlled by the same constraints.

\subsection{Ion-ion correlations}

\noindent Short-time cross-correlations between the mean displacements of the Cs$^+$ and TFSI$^-$ populations were computed using:
\begin{equation}
\label{eq:cross-correlation}
C_\text{Cs-TFSI} = \frac{\langle \mathbf{r}_\text{Cs}(t) \cdot 
\mathbf{r}_\text{TFSI}(t) \rangle}{\sqrt{\langle |\mathbf{r}_\text{Cs}(t)|^2 
\rangle \, \langle |\mathbf{r}_\text{TFSI}(t)|^2 \rangle}},
\end{equation}
with $\mathbf{r}_\alpha(t) = \mathbf{R}_\alpha(t) - \mathbf{R}_\alpha(0)$ the center-of-mass displacement of ionic species $\alpha$ over time $t$. In the limit $t \to 0$, $C_\text{Cs-TFSI}$ measures the degree of dynamical coupling of the instantaneous motions of the two ionic populations.

Our results show that $C_\text{Cs-TFSI}$ is largest at small $n$, indicating a tendency for Cs$^+$ and TFSI$^-$ to move in a concerted manner on short timescales (Fig.~\ref{supp-fig:cross-correlation}). As $n$ increases, $C_\text{Cs-TFSI}$ decreases systematically, reflecting progressively more independent ion motion. This trend is consistent with the enhanced ion clustering observed at small $n$ (Fig.~\ref{supp-fig:clusters}), as the motion of ions forming larger clusters is necessarily more correlated. This picture is also consistent with the (long-time) diffusion coefficients extracted from PFG-NMR and MSD analysis, suggesting that ion-ion dynamical coupling observed at short times may influence transport in the Fickian regime, although a quantitative connection between the two timescales remains to be established.

\subsection{Ionic conductivity}

\noindent The ionic conductivity $\sigma$ was measured experimentally. 
Our results indicate that $\sigma$ decreases from 
$\approx 2 \cdot 10^4$~\si{\micro\siemens\per\centi\meter} for the 
shortest chains ($n = 1$) to $\approx 3 \cdot 10^2$~\si{\micro\siemens\per\centi\meter} 
for the longest chains ($n = 88$) (Fig.~\ref{fig:sigma}A). Notably, 
the measured conductivity shows little dependence on the nature of the 
cation (Li$^+$, Na$^+$, Cs$^+$). The conductivity obtained from 
simulations (see details in the SI) displays a similar trend; however, the values are 
systematically lower, consistent with the lower diffusion coefficients 
reported previously.

\subsection{Degree of dissociation}

\noindent The degree of dissociation, $f_\mathrm{free}$, is estimated from the ratio between the measured ionic conductivity and the Nernst--Einstein conductivity,
\begin{equation}
f_\mathrm{free} \approx \frac{\sigma}{\sigma_\mathrm{NE}},
\end{equation}
with
\begin{equation}
\sigma_\mathrm{NE} = \frac{e^2}{k_\mathrm{B} T} \sum_i c_i z_i^2 D^i,
\end{equation}
where $e$ is the elementary charge, $k_\mathrm{B}$ is the Boltzmann constant, and $T$ is the temperature. The sum runs over all ionic species $i$, with $c_i$ the number concentration (in m$^{-3}$), $z_i$ their valency, and $D^i$ their self-diffusion coefficients. This expression assumes uncorrelated ionic motion, neglecting ion pairing 
and clustering, and therefore provides an upper bound to the conductivity.

For a symmetric 1:1 electrolyte, this reduces to
\begin{equation}
\sigma_\mathrm{NE} = \frac{F^2}{RT} \, C \left( D^+ + D^- \right),
\end{equation}
where $C$ is the salt concentration (in mol\,m$^{-3}$), $F = N_\mathrm{A} e$ is Faraday's constant, $N_\mathrm{A}$ the Avogadro number, and $R = N_\mathrm{A} k_\mathrm{B}$ is the gas constant, giving
\begin{equation}
\label{eq:ffree}
f_\mathrm{free} \approx \frac{\sigma RT}{F^2 \, C \left( D^+ + D^- \right)},
\end{equation}
where $D^+$ and $D^-$ are the diffusion coefficients of the alkali cation and TFSI$^-$ anion as measured by PFG-NMR.

\begin{figure}
\centering
\includegraphics[width=\columnwidth]{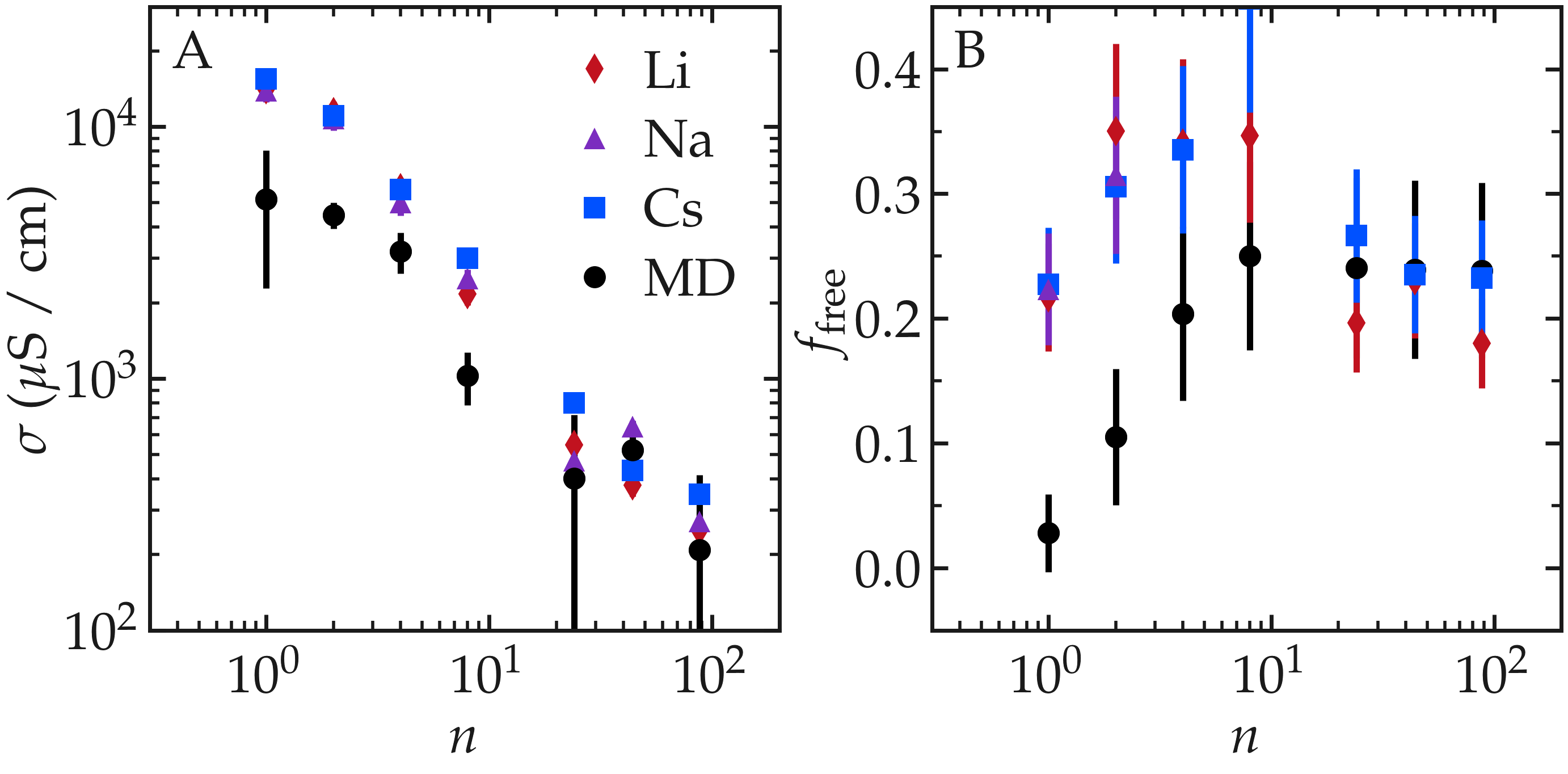}
\caption{A)~Ionic conductivity $\sigma$ as a function of the degree of polymerization, $n$. Results are from experiments, CsTFSI (blue squares), NaTFSI (magenta triangle), LiTFSI (red diamonds), and MD, CsTFSI (black disks). Error bars for MD represent standard deviations over multiple independent simulations. Experimental uncertainties are estimated to be $\approx 10~\%$. All measurements were performed at concentration $C = 0.8$~M and temperature $T = 60~^\circ\mathrm{C}$.
B)~Degree of dissociation, $f_\text{free}$, from experiments and MD as a function of polymer chain length $n$ [Eq.~\eqref{eq:ffree}]. Symbols and color coding are the same as in panel A. $f_\text{free}$ from MD is calculated based on cluster analysis, see text for details.}
\label{fig:sigma}
\end{figure}

The resulting $f_\mathrm{free}$ shows a non-monotonic dependence on chain length in experiments, with a maximum of $\approx 0.35$ near $n \approx 2$--$8$ (Fig.~\ref{fig:sigma}~B). No significant dependence on cation identity is observed. From MD, $f_\mathrm{free}$ can be estimated in two ways: via the Nernst--Einstein ratio [Eq.~\eqref{eq:ffree}], using $\sigma$ from non-equilibrium MD simulations and diffusion coefficients from MSD analysis; or from cluster analysis, where $f_\mathrm{free}$ is the fraction of 
Cs$^+$ ions not belonging to any ion cluster. Both methods yield consistent trends (Fig.~\ref{supp-fig:sigmaMD}), though cluster analysis gives slightly smaller values as it only accounts for unpaired Cs$^+$ and does not include TFSI$^-$ participation explicitly. In contrast with experiments, simulations predict a monotonic increase in $f_\mathrm{free}$ from $\approx 0.05$ to $\approx 0.25$ over the full range of chain lengths, with $f_\text{free}$ reaching a plateau for $n \ge 8$ (Fig.~\ref{fig:sigma}~B). 

\subsection{Cation transference number}

\noindent To further analyze the impact of the polymer chain length, $n$, on ion transport, we evaluated the cation transference number, $t^+$, which quantifies the relative contribution of cations to the overall ionic transport, defined as:
\begin{equation}
\label{eq:transference}
t^+ = \dfrac{D^+}{D^+ + D^-}.
\end{equation}
Our results indicate that the cation transference number, $t^+$, decreases with increasing $n$ (Fig.~\ref{supp-fig:transference}). $t^+$ remains close to 0.5 at small chain lengths, consistent with the similar cation and anion diffusivities observed for $n \lesssim 8$. It then decreases to approximately 0.3 at large $n$, reflecting the transition identified in the diffusion coefficients, where the anion becomes the more mobile species and contributes more strongly to charge transport than the cation. Transference numbers extracted from simulations are consistent with experiments, displaying the same trend (Fig.~\ref{supp-fig:transference}).

\section{Discussion}
\label{sec:discussion}

\subsection{Glyme diffusion scaling}

\noindent Our NMR results indicate that the scaling of glyme diffusion with chain length is largely insensitive to cation identity, despite substantial differences in ion coordination strength (Fig.~\ref{fig:diffusion}). Indeed, Li$^+$, owing to its high charge density, forms stronger coordination bonds with ether oxygens than the larger Cs$^+$, as reflected in binding energy measurements~\cite{memboeufStructureEnergeticsPolyEthylene2011}. This insensitivity to cation identity is consistent with prior simulations reporting little difference in conductivity between Na$^+$ and Li$^+$ electrolytes~\cite{wrobelNaFSINaTFSISolutions2021,liyana-arachchiPolarizableMolecularDynamics2018}. The slowing down of diffusion is thus dominated by the increasing size and relaxation times of the polymer chains as $n$ increases, rather than by specific ion--polymer interactions. Accordingly, glyme dynamics are primarily controlled by intrinsic segmental motion and available free volume, with ions reducing overall mobility (Fig.~\ref{supp-fig:diffusion-glymes}) without altering the underlying transport mechanism.

Deviations from ideal Rouse-like behavior ($D \propto n^{-1}$) indicate that free-volume constraints and local packing play an important role. Similar deviations were reported by Nam et al.~\cite{namDynamicsUnentangledCyclic2008} for dimethoxy-terminated poly(oxyethylene) at 56$~^\circ$C, where a stronger scaling exponent ($\alpha = -2.01$) was attributed primarily to free-volume effects. Additional contributions from proximity to the glass-transition temperature $T_\mathrm{g}$ may also play a role, since $T_\mathrm{g}$ itself depends on chain length through free-volume contributions of end groups~\cite{devauxMechanismIonTransport2012}.

\subsection{Structural origin of the crossover}
 
\noindent The crossover region near $n \approx 8$ may have a structural origin rooted in the coordination chemistry of the cations. Crystal structures of Cs$^+$ complexes with crown ethers and related cyclic polyethers indicate that approximately 6 ether oxygen atoms are required to complete the first coordination shell of Cs$^+$~\cite{ozutsumiXRayDiffractionStudy1989}. A linear glyme chain with $n \gtrsim 8$ therefore provides a sufficient number of ether oxygens to wrap around and fully coordinate a single cation. For shorter chains ($n \lesssim 8$), no single glyme molecule can provide full coordination: the cation must simultaneously coordinate multiple shorter chains, promoting the formation of multi-chain complexes. Similar crossover behavior has been reported in oligo(ethylene oxide)-based mesogenic systems, where the relative self-diffusion of Li$^+$, anions, and the host molecules depends strongly on chain length, with a transition from coupled ion-solvent motion at short polymer lengths to partial dynamical decoupling at longer lengths~\cite{judeinsteinIonicConductivityLithium2005}.
 
This picture is consistent with our MD simulations. For short chains, Cs$^+$ ions are typically coordinated by two or three distinct polymer chains simultaneously, and the first coordination shell contains a significant contribution from TFSI$^-$ oxygen atoms ($n_\mathrm{O\text{-}TFSI} \simeq 3.5$, Fig.~\ref{supp-fig:environments_Cs}). As chain length increases above $n \approx 8$, the number of coordinating chains collapses toward one, and anion participation in the solvation shell is progressively suppressed ($n_\mathrm{O\text{-}TFSI} < 2$, Fig.~\ref{supp-fig:environments_Cs_2}). Consistent with previous Raman spectroscopy studies of LiTFSI in glymes~\cite{brouilletteStableSolvatesSolution2002}, longer chains show a continuous increase in cation solvation that suppresses direct ion-pairing across the full concentration range. The coordination number $n_\mathrm{O\text{-}PEO}$ rises steadily with $n$ (Fig.~\ref{supp-fig:environments_Cs}), consistent with trends reported for Li$^+$ and Na$^+$ in oligoglymes~\cite{tangGlymesVersatileSolvents2014,
hendersonGlymeLithiumBistrifluoromethanesulfonylimideGlymeLithium2005}, where chain length was identified as a controlling variable for ion pairing, beyond bulk solvent polarity~\cite{plewa-marczewskaNMRStudiesEquilibriums2010}.

\subsection{Two transport regimes across the crossover}
 
\noindent A consistent crossover near $n \approx 8$ is observed in both PFG-NMR and MD simulations, reflecting a change in the dominant transport mechanism. For short chains ($n \lesssim 8$), glymes behave as low-viscosity molecular solvents. Ion--polymer residence times exceed characteristic diffusion times and polymer relaxation time ($\tau_\mathrm{res} \gg \tau_\alpha$, Fig.~\ref{fig:times}), indicating long-lived coordination complexes. Ions migrate together with their solvation shell in a vehicular mechanism, and cation and anion diffusivities are nearly equal, consistent with ion pairing. Short-time cross-correlations between Cs$^+$ and TFSI$^-$ are pronounced in this regime (Fig.~\ref{supp-fig:cross-correlation}), and ion clustering is enhanced (Fig.~\ref{supp-fig:clusters}). It should be noted that the measured glyme diffusion coefficient in this regime likely reflects a weighted average between free glyme molecules and glymes coordinated to ions, which diffuse more slowly as part of ion--solvent complexes. This picture is consistent with experimental observations in short glymes ($n \le 4$), where $D(\mathrm{Na}^+) \approx D(\mathrm{anion}) < D(\mathrm{solvent})$~\cite{carboneCharacteristicsGlymeElectrolytes2017,moralesIonTransportAssociation2019}. We note, however, that Park et al. recently reported that correlated ion motion has a reduced negative effect on conductivity at short chain lengths~\cite{parkMolecularDynamicsBasedOptimization2025}, a trend apparently opposite to ours. This discrepancy could originate from differences in salt concentration which can significantly alter the competition between ion--glyme and ion--ion interactions.
 
For longer chains ($n \gtrsim 8$), the ordering reverses: $\tau_\mathrm{res} \ll \tau_\alpha$. Transport becomes increasingly governed by local recoordination events rather than polymer motion, with the contribution of hopping-like transport growing in importance relative to vehicular motion. Ion--ion correlations weaken (Fig.~\ref{supp-fig:cross-correlation}) and clustering decreases (Fig.~\ref{supp-fig:clusters}), consistent with a more dynamically independent ionic environment. This picture is supported by simulations of Li$^+$ in PEO chains~\cite{diddensUnderstandingLithiumTransport2010}, which show that ion transport in long-chain systems is governed by intrachain and interchain coordination exchanges rather than polymer center-of-mass displacement, and is consistent with the general framework for ion transport in polymer electrolytes~\cite{meyerPolymerElectrolytesLithiumIon1998}. Mechanistic decomposition studies of PEO-LiTFSI at $n = 100$~\cite{leonMechanisticDecompositionIon2025} further show that rare cage-disassembly (hopping) events contribute disproportionately per event to the diffusion coefficient, supporting the efficiency of hopping-like transport. However, it has been shown that vehicular contributions are not negligible even at $n = 54$--$100$~\cite{borodinMechanismIonTransport2006,leonMechanisticDecompositionIon2025}, indicating that full decoupling from polymer dynamics is not achieved and that the transition is progressive rather than abrupt.

\subsection{Cation vs. anion mobility}
 
\noindent In PEO-based electrolytes, it is well established that anions tend to diffuse faster than cations, as the latter interact strongly with ether oxygens along the polymer backbone~\cite{goreckiPhysicalPropertiesSolid1995,hayamizu1Li7F192002,france-lanordEffectChemicalVariations2020}. Our results for $n \gtrsim 8$ are consistent with this picture, and help understand when and why this asymmetry develops.
 
At large $n$, TFSI$^-$ exhibits higher mobility than Cs$^+$ in both MD and PFG-NMR (Fig.~\ref{fig:diffusion}, Fig.~\ref{supp-fig:diffusion_coeff}). We attribute this asymmetry to differences in polymer interactions: Cs$^+$ remains partially coordinated to ether oxygens along the polymer backbone even in the long-chain regime, constraining its motion and slowing diffusion, while TFSI$^-$ interacts more weakly with the polymer matrix and diffuses more freely. This asymmetry grows with chain length: as the polymer backbone slows down and hopping-like transport grows in importance, the cation -- which must repeatedly exchange coordination sites along the slowly relaxing polymer -- is penalized more than the anion. As a result, the cation transference number decreases from $\approx 0.5$ at small $n$ toward $\approx 0.3$ at large $n$ (Fig.~\ref{supp-fig:transference}). By analogy with Li$^+$, for which quantum chemistry calculations show that longer glymes form more stable cation complexes while binding TFSI$^-$ more weakly~\cite{tsuzukiIntermolecularInteractionsLi2013}, a similar though likely weaker effect is expected for Cs$^+$ given its lower charge density. Similar behavior has been reported in other polymer electrolyte systems~\cite{edmanAnalysisDiffusionSolid2002}, confirming that $D(\mathrm{TFSI}^-) > D(\mathrm{cation})$ is a robust feature of the large-$n$ PEO regime.

\subsection{Ion dissociation and electrochemical implications}
 
\noindent The degree of ion dissociation, $f_\mathrm{free}$, estimated from the ratio of measured conductivity to the Nernst--Einstein conductivity [Eq.~\eqref{eq:ffree}], shows a non-monotonic dependence on chain length in experiments, with a maximum of $\approx 0.35$ near $n \approx 2$--$8$ (Fig.~\ref{fig:sigma}B). This maximum reflects the structural transition described above: at short $n$, cations must share their coordination among multiple chains and TFSI$^-$ anions, promoting ion pairing and cluster formation that suppress effective charge transport. As $n$ increases toward 8, each chain can fully coordinate a single cation, suppressing contact ion pairs and maximizing free-ion population. In contrast with experiments, simulations predict a continued increase in $f_\mathrm{free}$ with $n$ toward a plateau for $n \ge 8$, a discrepancy discussed below.
 
The chain-length dependence of ion transport thus reveals a trade-off between ion dissociation and cation transference. At short chain lengths ($n \lesssim 8$), ion pairing is detrimental to charge transport, as paired ions migrate as neutral units that contribute no ionic current. At long chain lengths ($n \gtrsim 8$), ion dissociation improves and clustering is reduced (Fig.~\ref{supp-fig:clusters}), but this comes at the cost of increasing anion dominance: the faster TFSI$^-$ carries a larger fraction of the charge, and the transference number declines toward $\approx 0.3$ (Fig.~\ref{supp-fig:transference}). A low transference number leads to concentration polarization and limits power density in battery applications~\cite{hillerInfluenceInterfacePolarization2013}. The system therefore presents unfavorable transport behavior at both extremes. The intermediate regime ($n \approx 2$--$8$) represents the best compromise, where $f_\mathrm{free}$ is maximized while the transference number has not yet declined substantially. These results motivate the search for strategies that decouple ion dissociation from the cation/anion mobility asymmetry~\cite{crabbElectrolyteDependenceLi2024}, such as anion immobilization~\cite{xuElectrolytesInterphasesLiIon2014} and single-ion conductors~\cite{bouchetSingleionBABTriblock2013}.

\subsection{Comparison between MD and PFG-NMR}
\label{sec:md-vs-nmr}
 
\noindent Overall, MD simulations reproduce the main experimental trends, including the scaling of diffusion coefficients with chain length and the crossover region near $n \approx 8$. However, quantitative discrepancies are observed: MD systematically underestimates diffusion coefficients and predicts slightly stronger scaling exponents. These differences can be attributed to several factors. Classical force fields neglect electronic polarization effects, which are known to improve the description of glyme-based electrolytes~\cite{liyana-arachchiPolarizableMolecularDynamics2018} but are computationally prohibitive at the microsecond timescales required here. Finite-size effects may also contribute, as the limited simulation box can artificially constrain long-range ion--ion correlations and extended cluster formation.
 
A more severe discrepancy concerns the degree of dissociation: experimentally, $f_\mathrm{free}$ shows a non-monotonic dependence on chain length with a maximum near $n \approx 2$--$8$, whereas MD predicts a monotonic increase with $n$~(Fig.~\ref{fig:sigma}). The experimental maximum likely reflects the formation of stable solvate structures at short chain lengths that suppress contact ion pairs~\cite{brouilletteStableSolvatesSolution2002}, a subtlety that non-polarizable force fields may not capture with sufficient accuracy. Additionally, the finite polydispersity of the experimental glyme samples (PDI $= 1.03$--$1.09$ for the longest chains) means that a small fraction of shorter, more mobile chains is inevitably present. As shown by Thiam et al.~\cite{thiamPEOImmobileSolvent2019}, such unentangled oligomers move vehicularly with ions and can artificially enhance the measured diffusion coefficients, potentially contributing to the systematic overestimation of experimental values relative to MD at large $n$.
 
Despite these quantitative differences, MD simulations and PFG-NMR are complementary: PFG-NMR provides direct access to transport coefficients and reveals their non-trivial dependence on chain length, whereas MD offers molecular-level resolution of ion--polymer coordination, ion pairing, and dynamical heterogeneities. The combination of both approaches has been central to the molecular picture developed here.

\section{Conclusion}

\noindent We studied ion transport in glyme-based electrolytes over a wide chain-length range ($n = 1$--$88$) using PFG-NMR to examine Li$^+$, Na$^+$, and Cs$^+$ paired with TFSI$^-$. Our combined use of PFG-NMR, conductivity measurements, and MD simulations allows us to connect macroscopic transport coefficients to their microscopic origins. While PFG-NMR provides direct access to ion and glyme self-diffusion coefficients, and ionic conductivity probes the mobility of charged species, MD reveals the underlying mechanisms in terms of ion residence times, coordination dynamics, and ion-ion correlations. Glyme diffusion scales as $D \propto n^{-1.5}$, independent of the cation, indicating that free-volume and packing constraints dominate polymer dynamics. A crossover region at $n \approx 8$ separates two regimes. For short chains ($n \lesssim 8$), transport is vehicular: ions remain bound to their solvation shell, ion pairing is strong, and cation/anion diffusion is similar. For long chains ($n \gtrsim 8$), ion transport decouples from polymer motion via rapid coordination exchange, with weaker ion pairing and improved dissociation. Notably, increasing chain length enhances anion mobility relative to cations, reducing the transference number from $\approx 0.5$ to $\approx 0.3$. Beyond glyme systems, these results help establish a general physical framework for ion transport in coordinating polymer electrolytes, where transport is controlled by the competition between polymer relaxation and ion coordination dynamics.

\section*{Supplementary information}

\noindent Additional details supporting the analysis are provided in the Supplementary Information. This includes the procedure for identifying ion clusters and extracting the degree of dissociation, and the analysis of cation residence times on PEO chains. We also detail the calculation of the segmental (orientational) relaxation times of the polymer, as well as the protocol used to compute ionic conductivity from non-equilibrium molecular dynamics simulations. In addition, the SI contains all simulation compositions, force-field parameters, and the complete set of supplementary tables and figures referenced in the main text.

\section*{Acknowledgement}

\noindent This work was supported by a public grant from the “Laboratoire d'Excellence Physics Atoms Light Mater” (LabEx PALM), overseen by the French National Research Agency (ANR) as part of the “Investissements d'Avenir” program (reference: ANR-10-LABX-0039-PALM). S.G. also acknowledge funding from the ANR under grant ANR-24-CE06-5671 (MicroSep). Some of the computations presented in this paper were performed using the GRICAD infrastructure (\href{https://gricad.univ-grenoble-alpes.fr}{gricad.univ-grenoble-alpes.fr}), which is supported by Grenoble research communities. In addition, this work was granted access to the HPC resources of IDRIS under the allocation 2025-A0192A14560 made by GENCI. Claire Goldmann is acknowledged for assistance with chemistry and sample preparation. Sylvain Franger and Benjamin Rondeau from ICMMO are acknowledged for providing access to ionic conductivity measurements. CNRS, CEA, and Université Paris-Saclay are acknowledged for their recurrent funding. This paper is dedicated to our late colleague and friend Mehdi Zeghal, who was a key driving force behind the conception, execution, and interpretation of this work.

\bibliographystyle{ieeetr}
\bibliography{zotero-export,additional}

\end{document}


\title{Supplementary Information: \\ Molecular-to-polymeric crossover in ion diffusion in glyme-based electrolytes: \\
from vehicular to hopping transport}

\author{Aicha Jani}

\affiliation{Université Paris-Saclay, CNRS, LPS, 91405 Orsay, France}

\author{Simon Gravelle}
\email{simon.gravelle@cnrs.fr}

\affiliation{Université Grenoble Alpes, CNRS, LIPhy, 38000 Grenoble, France}

\author{Pawel Wzietek}
\affiliation{Université Paris-Saclay, CNRS, LPS, 91405 Orsay, France}

\author{Mehdi Zeghal$^{\dagger}$}

\affiliation{Université Paris-Saclay, CNRS, LPS, 91405 Orsay, France}

\author{Patrick Judeinstein}

\affiliation{Université Paris-Saclay, CNRS, LPS, 91405 Orsay, France}
\affiliation{Université Paris-Saclay, CNRS, CEA, LLB, 91191 Gif-sur-Yvette, France}

\maketitle

$\dagger$ in memoriam of our colleague Mehdi Zeghal

\section{Supplementary methods}

\subsection{Cluster analysis}

\begin{figure}
\centering
\includegraphics[width=0.7\columnwidth]{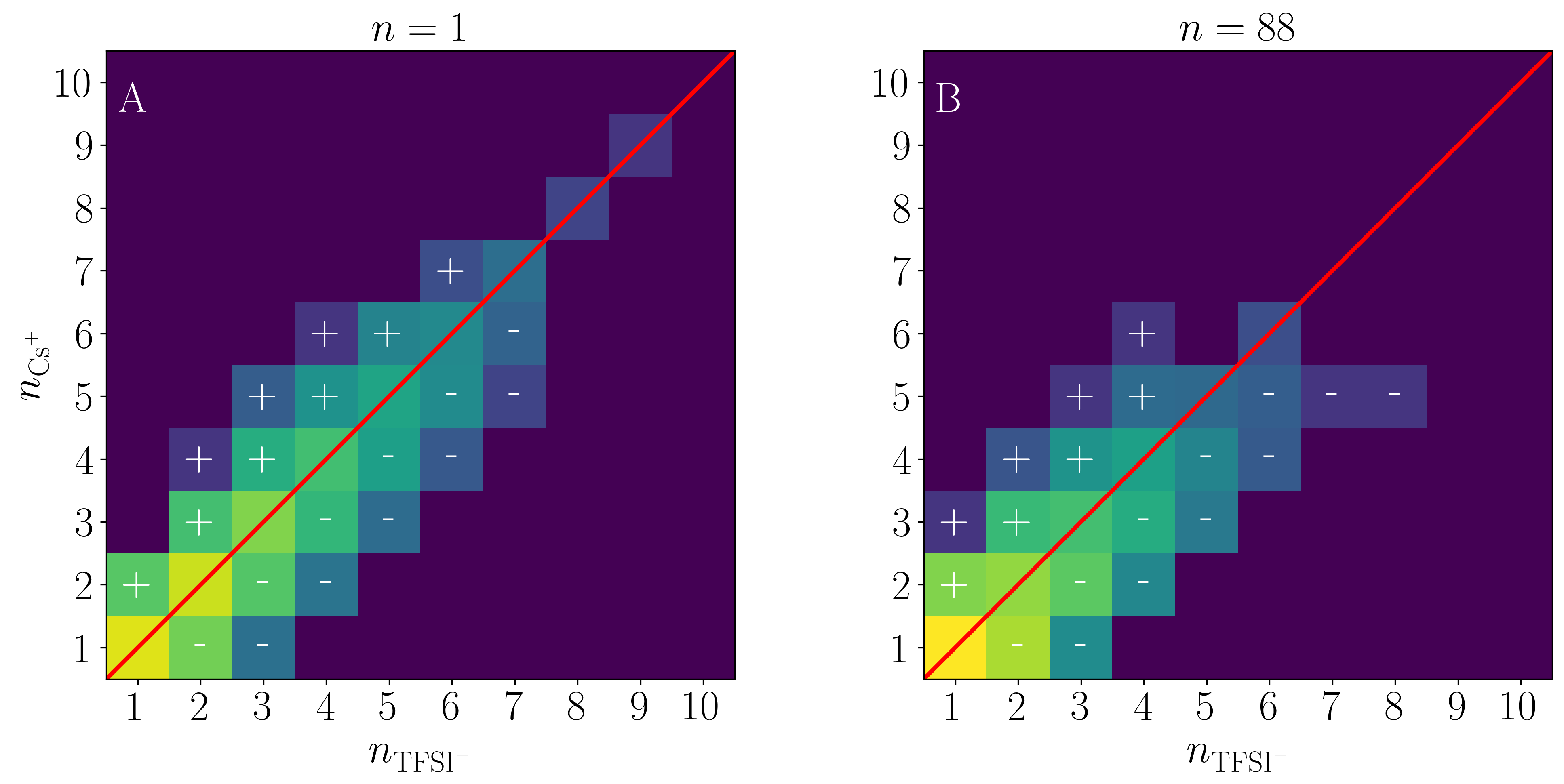}
\caption{
Cluster matrices showing the probability distribution of Cs$^+$ cations and TFSI$^-$ anions in a cluster for $n=1$ (A) and $n = 88$ (B).
Yellow indicates high probability, while blue corresponds to low or zero probability.
Symbols within each cell indicate the sign of the cluster's net charge. 
The red diagonal line passes through cells with equal numbers of Cs$^+$ and TFSI$^-$ ions, indicating balanced clusters
(conditions: $C = 0.8$~M, $T = 60~^\circ\mathrm{C}$).
}
\label{fig:clusters}
\end{figure}

\begin{figure}
\centering
\includegraphics[width=0.3\columnwidth]{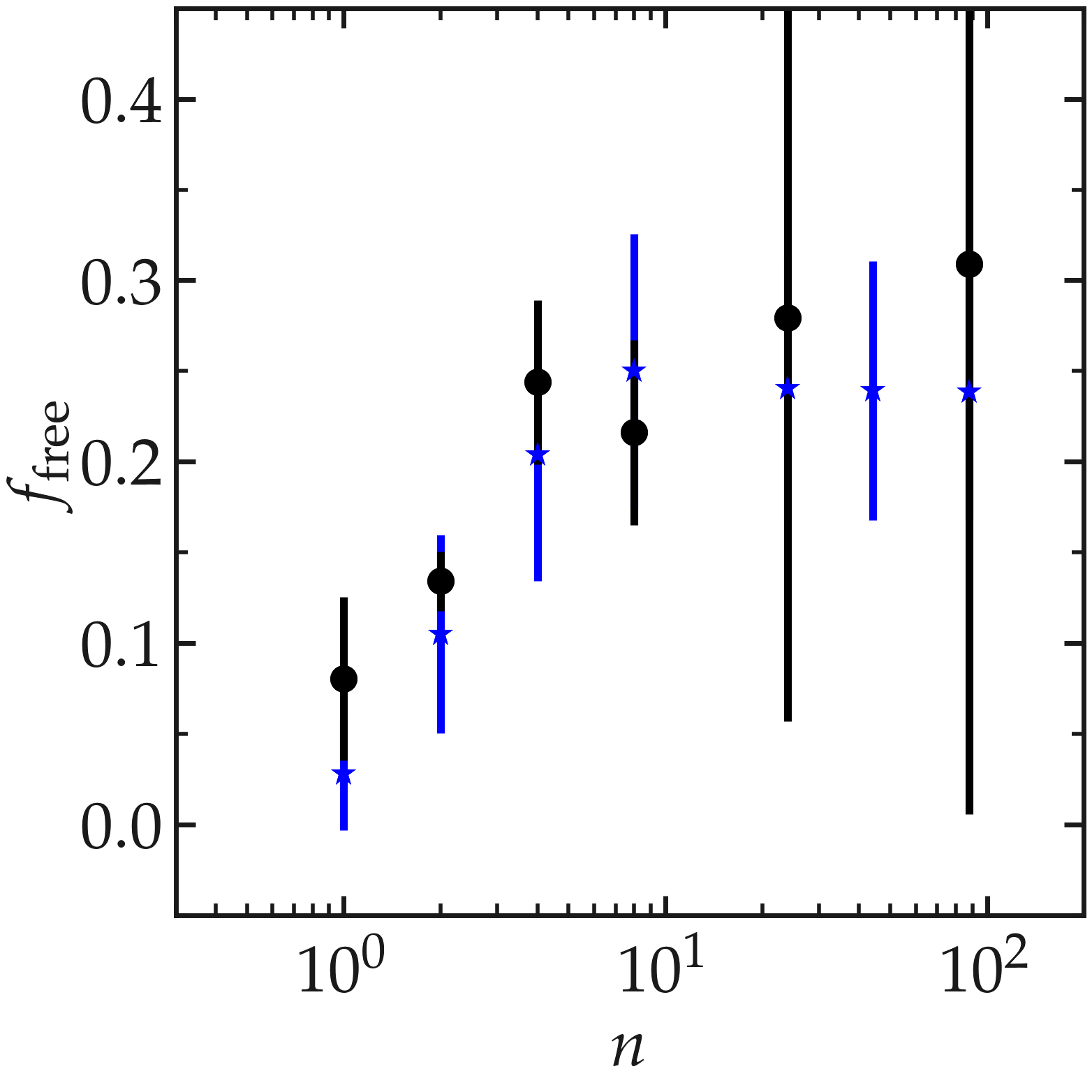}
\caption{Degree of dissociation $f_\text{free}$ from MD simulations 
as a function of polymer chain length $n$, estimated from the 
Nernst--Einstein ratio [Eq.~\eqref{main-eq:ffree} from the main text] (black disks) 
and from cluster analysis (blue stars) (conditions: $C = 0.8$~M, $T = 60~^\circ\mathrm{C}$).}
\label{fig:sigmaMD}
\end{figure}

\noindent Ion clusters were identified from MD trajectories using a distance-based criterion. At each frame, Cs$^+$ and TFSI$^-$ ions were considered in contact if the distance between the cation and any TFSI oxygen atom is below a cutoff $r_\text{c} = 3.9$~\AA{}, determined from the first minimum of the cation--oxygen radial distribution function. Clusters were built by grouping all ions connected through a chain of such contacts. For each cluster, the number of Cs$^+$ and TFSI$^-$ ions was recorded, and the cluster size distribution was accumulated over the trajectory (Fig.~\ref{fig:clusters}). From the cluster analysis, the degree of dissociation,  $f_\mathrm{free}$, was also estimated from the fraction of cations not belonging to any cluster, showing good agreement with $f_\mathrm{free}$ measured from Eq.~\eqref{main-eq:ffree} of the main text (Fig.~\ref{fig:sigmaMD}).

\subsection{Cation residence times}

\noindent The average residence time $\tau_\mathrm{res}$ of a cation on a PEO chain was computed using a first-passage analysis applied to cation--chain contacts (Fig.~\ref{main-fig:times} of the main text). A cation is considered coordinated to a given PEO chain if at least one oxygen atom of that chain lies within $r_\text{c} = 4.3$~\AA{} of the cation, as determined from the first minimum of the cation--oxygen radial distribution function. A binary contact variable $m(t)$ was constructed for each cation--chain pair, with $m = 1$ when the cation is coordinated to that chain and $m = 0$ otherwise. 

From these binary time series, the first-passage probability distributions $\Psi_\text{bound}(t)$ and $\Psi_\text{unbound}(t)$ were computed, where $\Psi_\text{bound}(t)$ ($\Psi_\text{unbound}(t)$) is the distribution of times spent in the bound (unbound) state before the first transition to the unbound (bound) state. The characteristic residence time $\tau_\mathrm{res}$ is defined as the mean of $\Psi_\text{bound}(t)$, and all quantities are averaged over all cation--chain pairs in the simulation box.

\subsection{Orientational relaxation time of the PEO}

\noindent Segmental dynamics of the PEO chains were characterized from the autocorrelation of local bond vectors: First, the normalized bond vector $\mathbf{u}(t)$ connecting two ether oxygen atoms along the polymer backbone (spanning one $-\mathrm{CH_2{-}CH_2}-$ unit) was computed. Then, the orientational relaxation was quantified through the time autocorrelation function
\begin{equation}
\label{eq:bond-vector}
C(t) = \left\langle \mathbf{u}(0)\cdot\mathbf{u}(t) \right\rangle ,
\end{equation}
evaluated over the trajectory. The decay of $C(t)$ reflects the progressive loss of orientational memory associated with local polymer motions.

For each system, the procedure was repeated for multiple randomly chosen chains and neighboring bond pairs, yielding a distribution of correlation functions and relaxation times. A characteristic relaxation time $\tau_\alpha$ was extracted from each trajectory as the time at which the autocorrelation function first decays to $1/e$. The resulting distributions of $\tau_\alpha$ were then averaged to obtain the mean segmental relaxation time and its standard deviation.

\subsection{Ionic conductivity from non-equilibrium MD}

\noindent The ionic conductivity $\sigma$ was computed from non-equilibrium MD (NEMD) simulations using the linear response to a small applied electric field. A static electric field of amplitude $E = 0.01$~V\,nm$^{-1}$ was applied along one of the axis, inducing a steady-state ionic current. The field amplitude was chosen small enough to remain in the linear response 
regime. All other simulation parameters were identical to the production runs described in the main text, with the exception that pressure coupling was disabled. Simulations were run for 100~ns.

The ionic current density was computed from the charge flux along the field direction,
\begin{equation}
J = \frac{1}{V} \sum_i \frac{q_i \, \Delta x_i}{\Delta t},
\end{equation}
where $q_i$ is the charge of ion $i$, $\Delta x_i$ is its displacement along $x$ between consecutive frames, $\Delta t$ is the frame interval, and $V$ is the simulation box volume. The sum runs over all Cs$^+$ and TFSI$^-$ ions. The conductivity was then obtained from Ohm's law,
\begin{equation}
\sigma = \frac{J}{E}.
\end{equation}
To estimate statistical uncertainty, each trajectory was divided into 10 equal blocks, $\sigma$ was computed independently for each block, and the reported value is the block mean with its standard error.

\clearpage

\section{Tables and figures}

\begin{table}[h!]
\centering
\begin{tabular}{lccc}
Glyme & $n$ & Melting point (\textdegree C) & Boiling point (\textdegree C) \\
MG & 1 & -58$^a$ & 85$^a$ \\
DG & 2 & -64$^a$ & 162$^a$ \\
4G & 4 & -27$^a$ & 275$^a$ \\
P400 & 8 & 10$^a$ & $\textgreater 300^a$ \\
P1100 & 24 & 36$^b$ & $\textgreater 300^b$ \\
P2000 & 44 & 50$^b$ & $\textgreater 300^b$ \\
P4000 & 88 & 59$^b$ & $\textgreater 300^b$ \\
\end{tabular}
\caption{Melting and boiling points of different glymes. Footnotes indicate source: $^a$~Merck, $^b$~Polymer Source.}
\label{tab:melting-glym}
\end{table}

\begin{table}[h!]
\centering
\begin{tabular}{ccccc}
$n$ & PEO chains & TFSI$^-$ & Cs$^+$ & Box size (nm) \\
1  & 376 & 25 & 25 & 4.3 \\
2  & 261 & 26 & 26 & 4.2 \\
4  & 163 & 27 & 27 & 4.2 \\
8  & 93  & 27 & 27 & 4.1 \\
24 & 34  & 28 & 28 & 4.1 \\
44 & 19  & 28 & 28 & 4.1 \\
88 & 10  & 29 & 29 & 4.1 \\
\end{tabular}
\caption{Composition of the simulated systems for different degrees of polymerization $n$. The table reports the number of PEO chains, TFSI$^-$ anions, and Cs$^+$ cations, along with the average edge length of the cubic simulation box.}
\label{tab:system_composition}
\end{table}

\begin{table}[h!]
\centering
\begin{tabular}{lcccc}
Atom type & Element & $q$ (e) & $\sigma$ (nm) & $\epsilon$ (kJ mol$^{-1}$) \\
\multicolumn{5}{c}{PEO} \\
CC32A  & C  & $-0.01$ & 0.35814 & 0.23430 \\
OC30A  & O  & $-0.34$ & 0.29399 & 0.41840 \\
HCA2   & H  & $+0.09$ & 0.23876 & 0.14644 \\
\multicolumn{5}{c}{TFSI$^-$} \\
Ci     & C  & $+0.35$ & 0.31500 & 0.08281 \\
Fli    & F  & $-0.16$ & 0.26550 & 0.06651 \\
Sui    & S  & $+1.02$ & 0.40825 & 0.31372 \\
Nii    & N  & $-0.66$ & 0.32500 & 0.21333 \\
Oi     & O  & $-0.53$ & 0.34632 & 0.26353 \\
\multicolumn{5}{c}{Cs$^+$} \\
Csi    & Cs & $+1.00$ & 0.51700 & 0.00061 \\
\end{tabular}
\caption{Non-bonded force field parameters used in molecular dynamics simulations. Lennard-Jones parameters ($\sigma$, $\epsilon$) and partial charges ($q$) are given for each atom type.}
\label{tab:ff_parameters}
\end{table}

\begin{table}[h!]
\centering
\begin{tabular}{cccccc}
$n$ & NMR (no salt) & MD (no salt) & NMR (CsTFSI) & MD (CsTFSI) \\
1  & 0.467 & 0.460 & 0.269 & 0.365 \\
2  & 0.206 & 0.180 & 0.117 & 0.128 \\
4  & 0.080 & 0.051 & 0.042 & 0.035 \\
8  & 0.020 & 0.012 & 0.012 & 0.009 \\
24 & $4.961 \times 10^{-3}$ & $1.446 \times 10^{-3}$ & $3.271 \times 10^{-3}$ & $9.98 \times 10^{-4}$ \\
44 & $1.623 \times 10^{-3}$ & $5.04 \times 10^{-4}$ & $1.126 \times 10^{-3}$ & $2.531 \times 10^{-4}$ \\
88 & $5.678 \times 10^{-4}$ & $2.137 \times 10^{-4}$ & $3.552 \times 10^{-4}$ & $9.234 \times 10^{-5}$ \\
\end{tabular}
\caption{Comparison between experimental (PFG-NMR) and simulation (MD) self-diffusion coefficients of glymes ($\times 10^{-8}$m$^2$s$^{-1}$). All measurements were performed at concentration $C = 0.8$~M and temperature $T = 60~^\circ\mathrm{C}$.}
\label{tab:label-D-glym}
\end{table}

\begin{figure*}
\centering
\includegraphics[width=0.65\columnwidth]{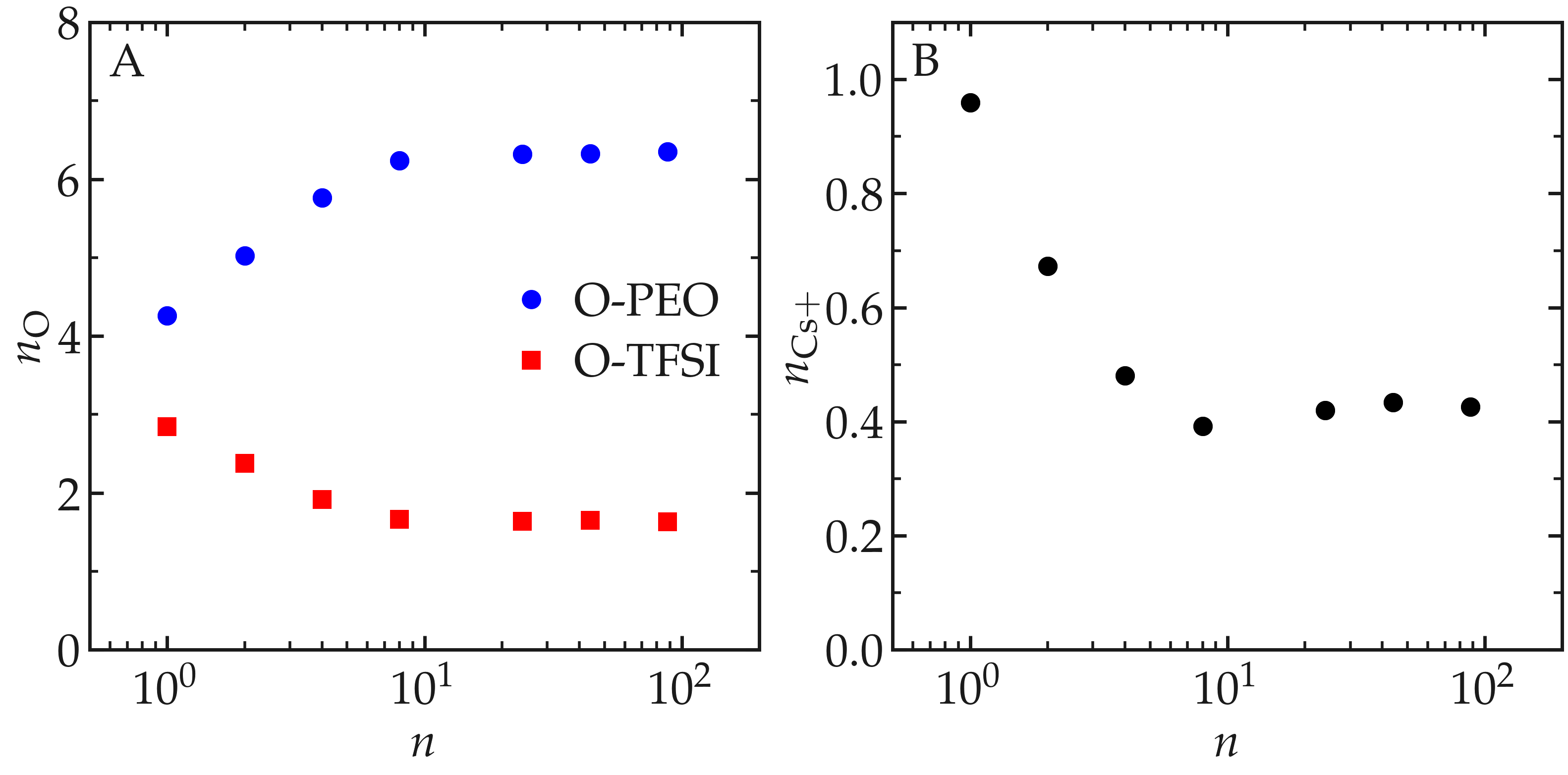}
\caption{(A) Average number of neighboring oxygen atoms within a cutoff distance $r_\text{c}$ of a Cs$^+$ ion as a function of the degree of polymerization, $n$, as measured from MD simulations. The cutoff distance is determined from the first minimum of the cation--oxygen radial distribution function. For oxygen atoms belonging to a PEO chain (blue disks), $r_\text{c} = 4.3~\mathrm{\AA}$, whereas for oxygen atoms from TFSI$^-$ (red squares), $r_\text{c} = 3.9~\mathrm{\AA}$. (B) Average number of neighboring Cs$^+$ ions within a cutoff distance $r_\text{c} = 8.5~\mathrm{\AA}$ of a Cs$^+$ ion, where the cutoff is determined from the first minimum of the cation--cation radial distribution function
(conditions: $C = 0.8$~M, $T = 60~^\circ\mathrm{C}$).}
\label{fig:environments_Cs}
\end{figure*}

\begin{figure*}
\centering
\includegraphics[width=\columnwidth]{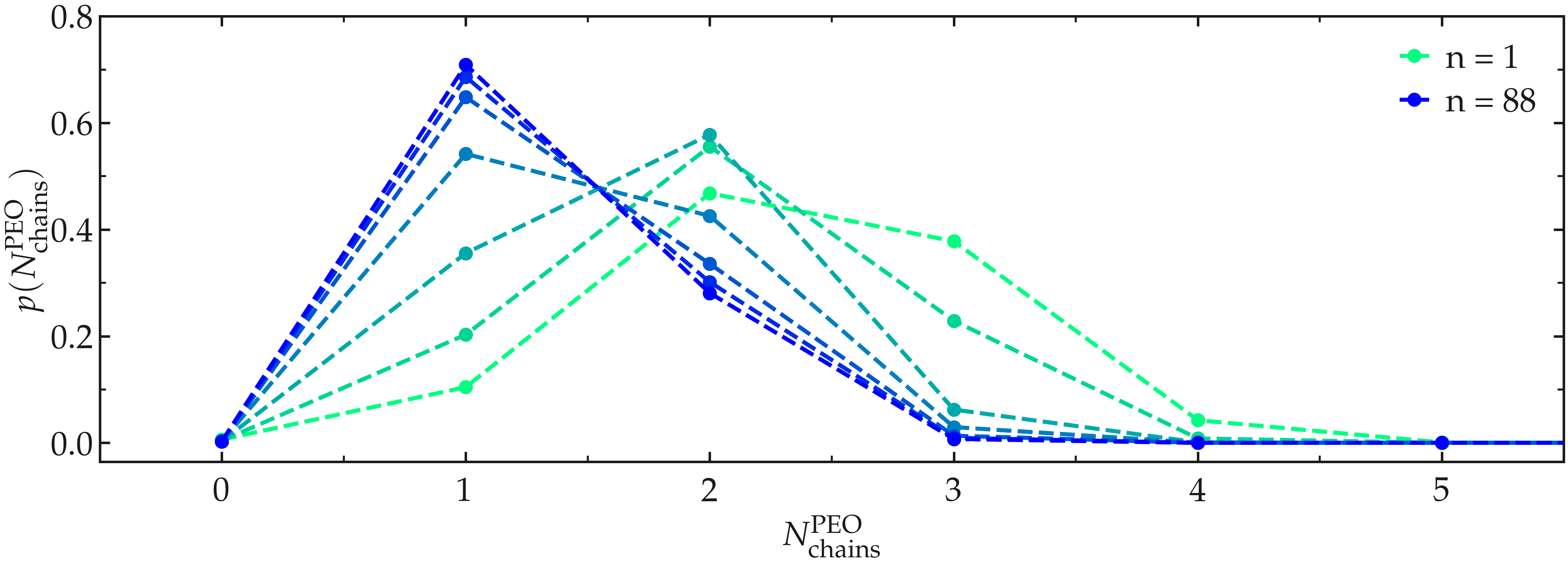}
\caption{Probability density distribution of the number of distinct PEO chains, $N_{\mathrm{chains}}^{\mathrm{PEO}}$, coordinating a Cs$^+$ ion for different degrees of polymerization, $n$, from green to blue: $n = 1$, $2$, $4$, $8$, $24$, $44$, and $88$. The distribution reflects the typical number of individual PEO chains participating in the solvation of a single cation (conditions: $C = 0.8$~M, $T = 60~^\circ\mathrm{C}$).}
\label{fig:environments_Cs_2}
\end{figure*}

\begin{figure}
\centering
\includegraphics[width=0.35\columnwidth]{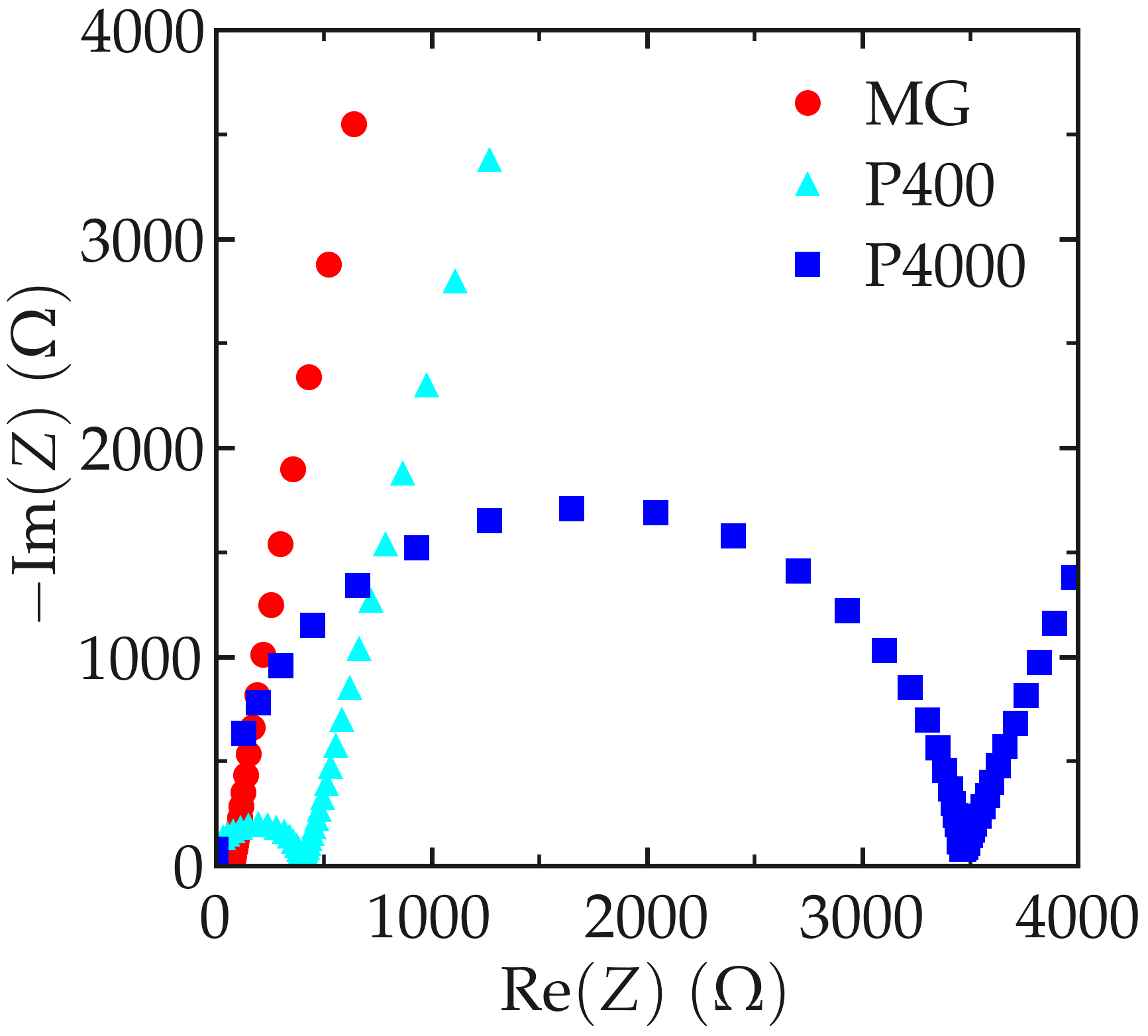}
\caption{
Nyquist plots obtained from electrochemical impedance spectroscopy measurements for MG (red disks), P400 (cyan triangles), and P4000 (blue squares) glymes containing NaTFSI. The real part of the impedance is plotted as a function of the negative imaginary part. The high-frequency intercept of each spectrum with the real axis corresponds to the bulk resistance $R$, which was used to calculate the ionic conductivity according to Eq.~\eqref{main-eq:impedance} from the main text (conditions: $C = 0.8$~M, $T = 60~^\circ\mathrm{C}$).
}
\label{fig:nyquist}
\end{figure}

\begin{figure}
\centering
\includegraphics[width=0.7\columnwidth]{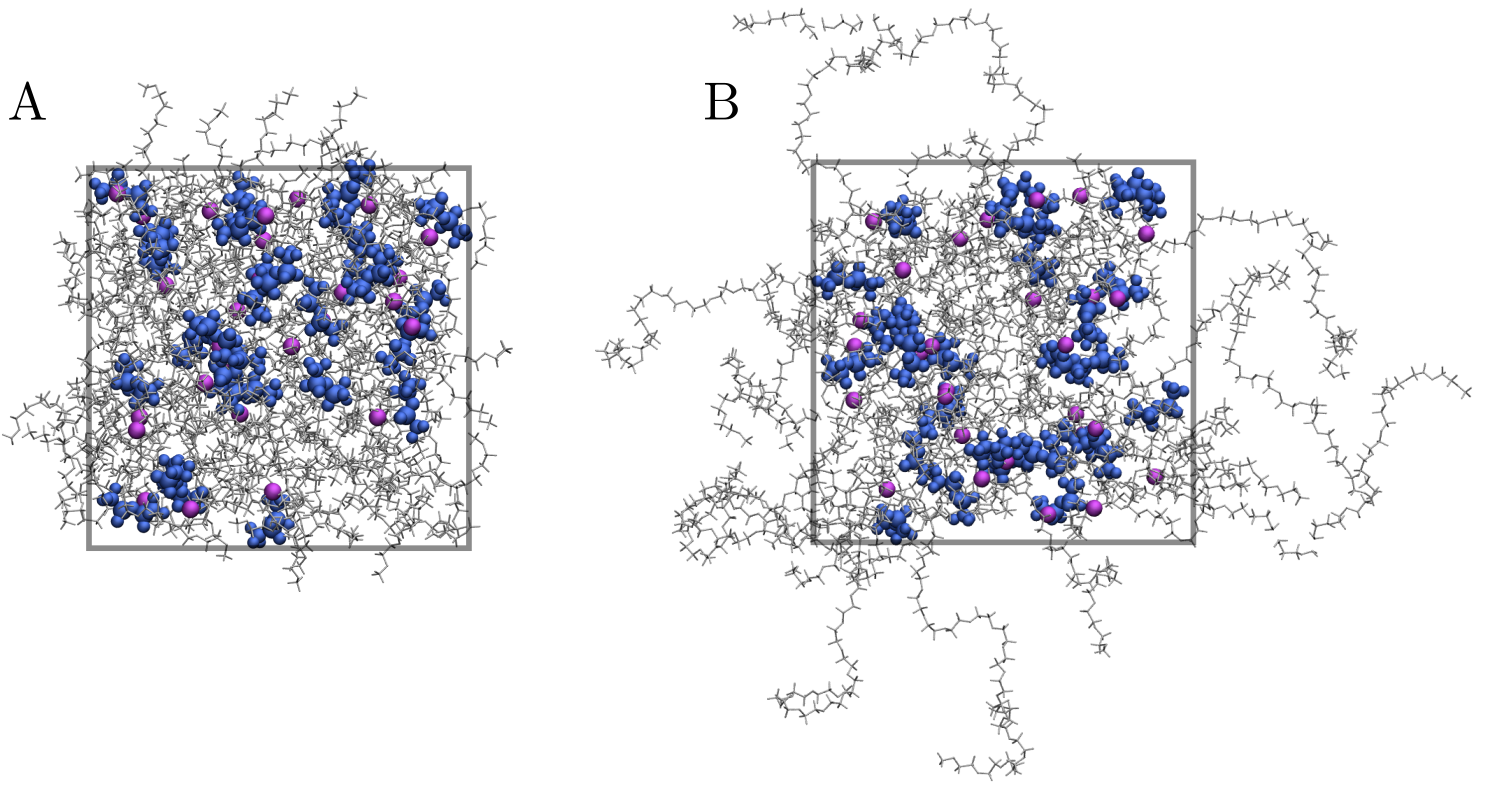}
\caption{Snapshots of systems from MD simulations. PEO chains are in gray, Cs$^+$ ions in purple, and TFSI$^-$ ions in blue. A)~Short PEO chains with $n = 4$. B)~Long PEO chains with $n = 24$.}
\label{fig:systems}
\end{figure}

\begin{figure}
\centering
\includegraphics[width=0.55\columnwidth]{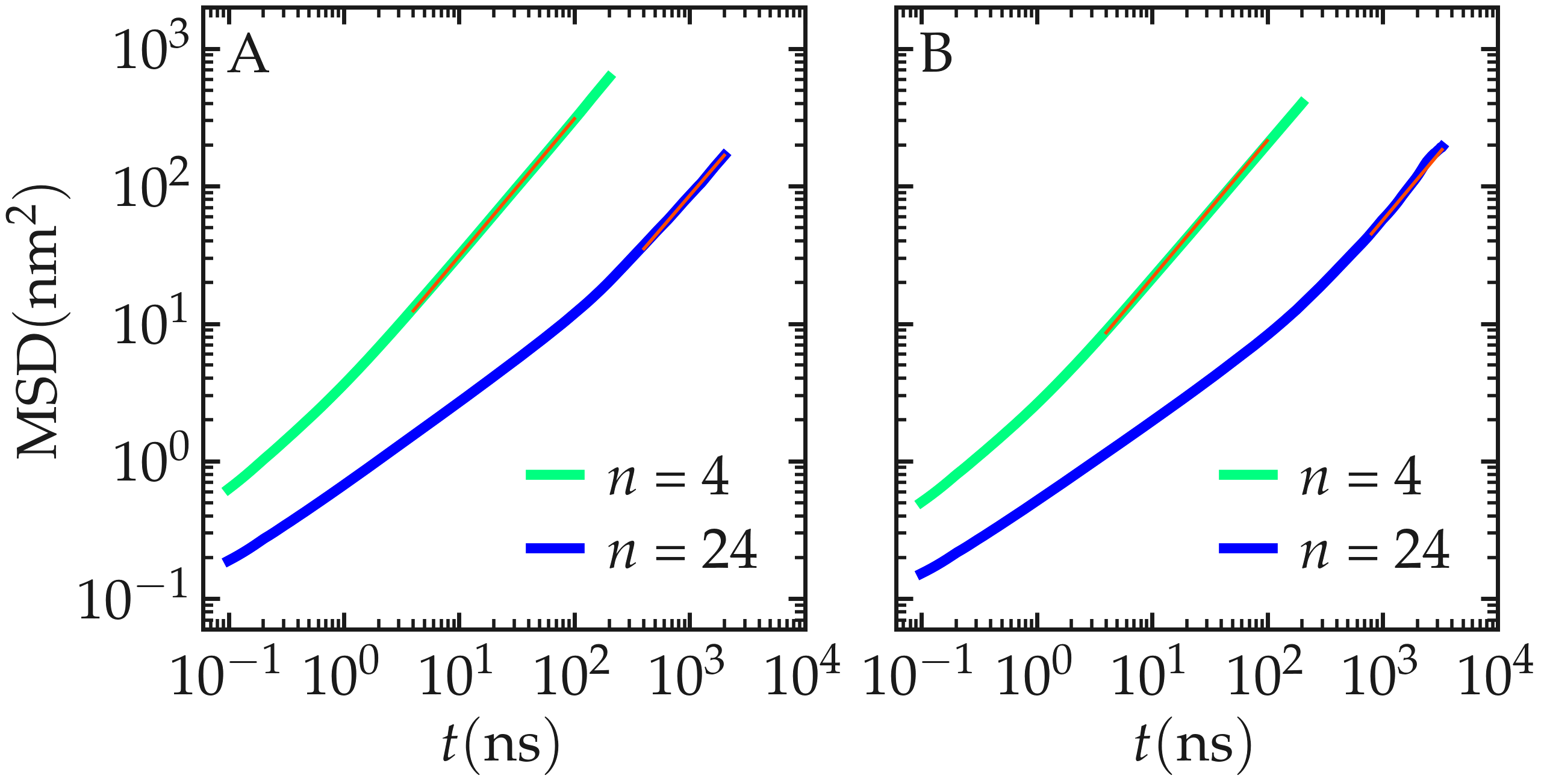}
\caption{A)~Mean-squared displacement (MSD) of glymes in the absence of salt from MD simulations for two values of $n$. The guiding lines indicate linear fits performed in the diffusive regime. B)~MSD of glymes with CsTFSI (conditions: $C = 0.8$~M, $T = 60~^\circ\mathrm{C}$).}
\label{fig:msd}
\end{figure}

\begin{figure}
\centering
\includegraphics[width=0.55\columnwidth]{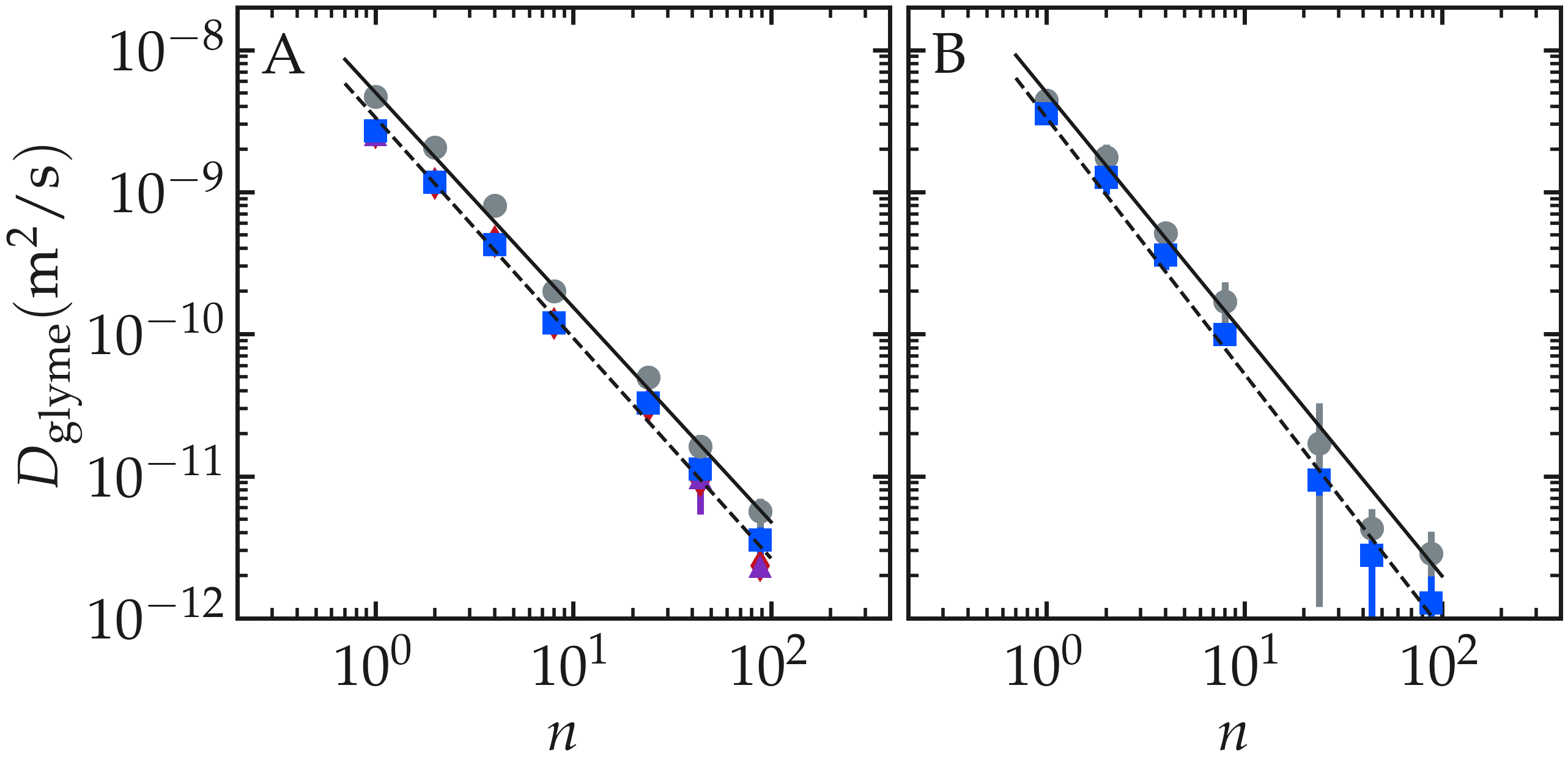}
\caption{A)~Self-diffusion coefficients for glymes, $D_\mathrm{glyme}$, from PFG-NMR without salt (gray disks) and with CsTFSI (blue squares), NaTFSI (magenta triangles), LiTFSI (red diamonds) as a function of glyme chain length $n$. Guiding lines indicate trends with slopes of -1.51 (solid line) and -1.55 (dashed line).
B)~$D_\mathrm{glyme}$ from MD for glyme without salt (gray disks) and with CsTFSI (blue squares). Guiding lines indicate trends with slopes of -1.7 (solid line) and -1.8 (dashed line)
(conditions: $C = 0.8$~M, $T = 60~^\circ\mathrm{C}$).
}
\label{fig:diffusion-glymes}
\end{figure}

\begin{figure}
\centering
\includegraphics[width=0.55\columnwidth]{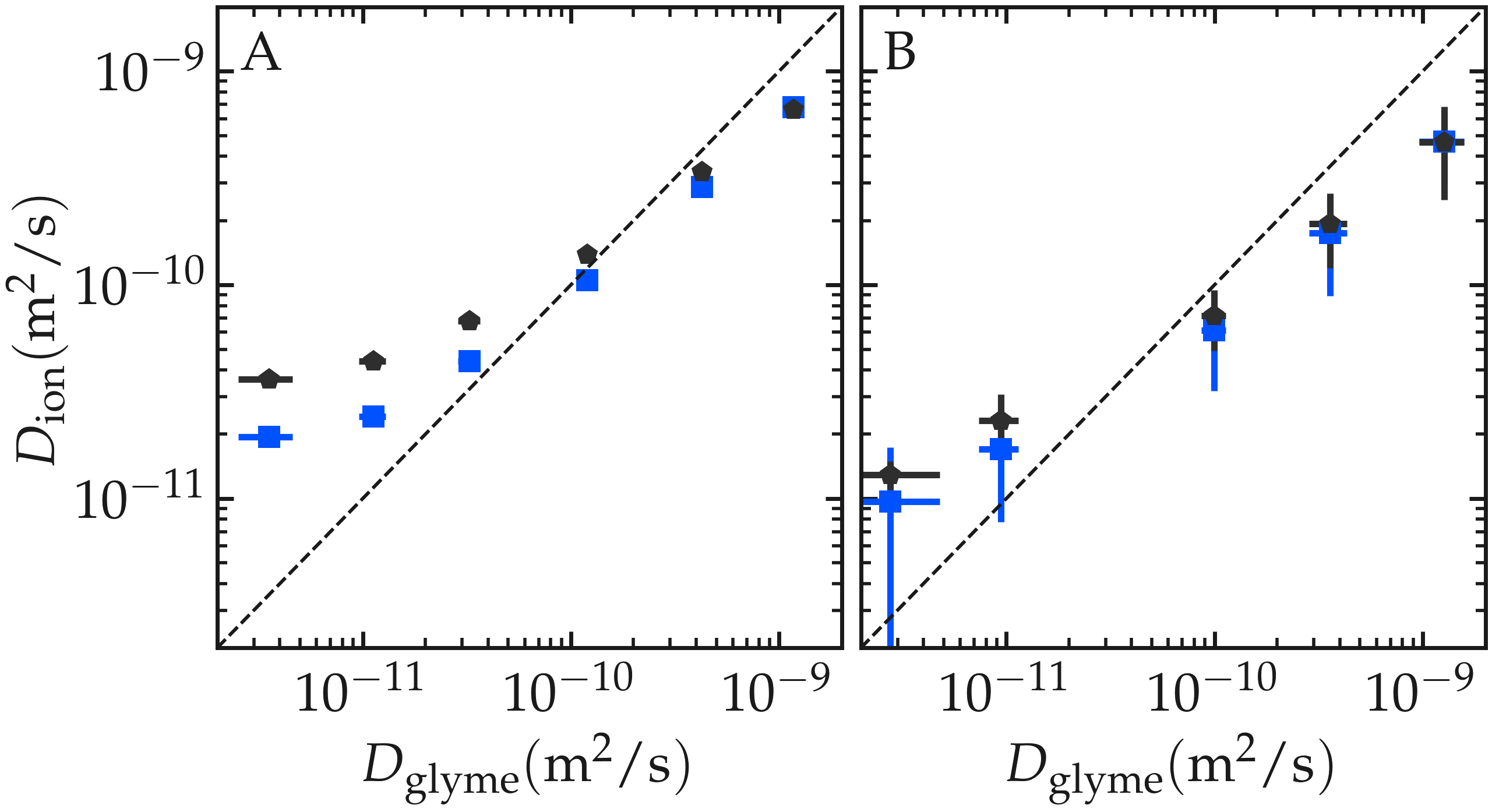}
\caption{Self-diffusion coefficients of Cs$^+$ (blue diamonds) and TFSI$^-$ (black stars) plotted as a function of the glyme self-diffusion coefficient. The dashed line indicates the reference condition $D_{\mathrm{ion}} = D_{\mathrm{glyme}}$. (A)~PFG-NMR measurements. (B)~Molecular dynamics simulations (conditions: $C = 0.8$~M, $T = 60\,^\circ\mathrm{C}$).}
\label{fig:diffusion_coeff}
\end{figure}

\begin{figure*}
\centering
\includegraphics[width=\columnwidth]{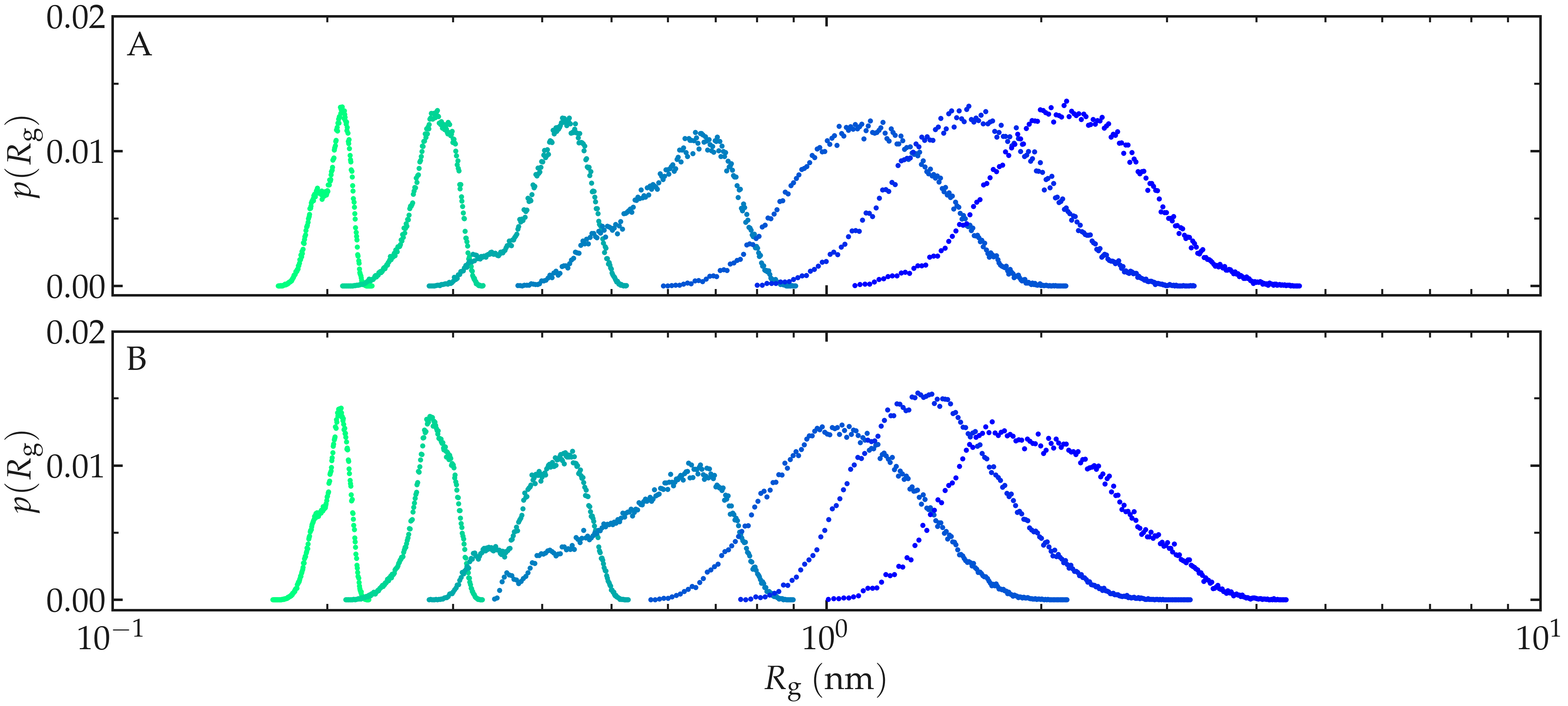}
\caption{
(A) Probability distribution of the radius of gyration $R_\text{g}$ of glymes obtained from MD simulations in the absence of salt. 
Chains have varying degrees of polymerization, from left to right: $n = 1$, $2$, $4$, $8$, $24$, $44$, and $88$.
(B) Probability distribution of the radius of gyration $R_\text{g}$ of glymes in the presence of CsTFSI
(conditions: $C = 0.8$~M, $T = 60~^\circ\mathrm{C}$).
}
\label{fig:gyration}
\end{figure*}

\begin{figure}
\centering
\includegraphics[width=0.7\columnwidth]{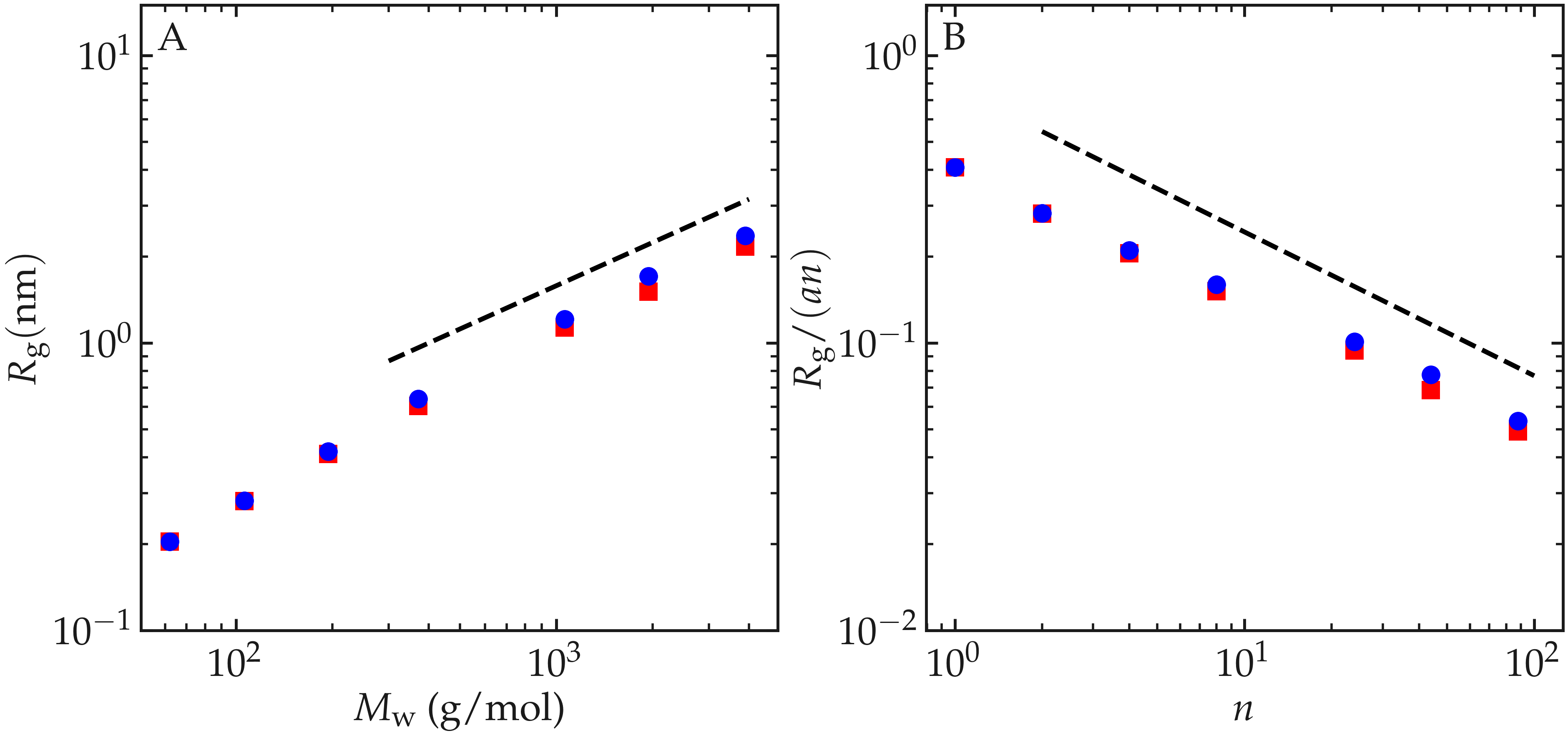}
\caption{
A)~Average radius of gyration, $R_\mathrm{g}$, as a function of the molar mass of the glymes, $M_\mathrm{w}$, measured from MD simulations in the absence (blue circles) and in the presence (red squares) of CsTFSI. The dashed line is $\propto M_\mathrm{w}^{0.5}$. B)~$R_\mathrm{g} / (a n)$, as a function of the degree of polymerization, $n$, with $a = 0.5~\text{nm}$. The dashed line is $\propto n^{-0.5}$
(conditions: $C = 0.8$~M, $T = 60~^\circ\mathrm{C}$).
}
\label{fig:gyration2}
\end{figure}

\begin{figure}
\centering
\includegraphics[width=0.3\columnwidth]{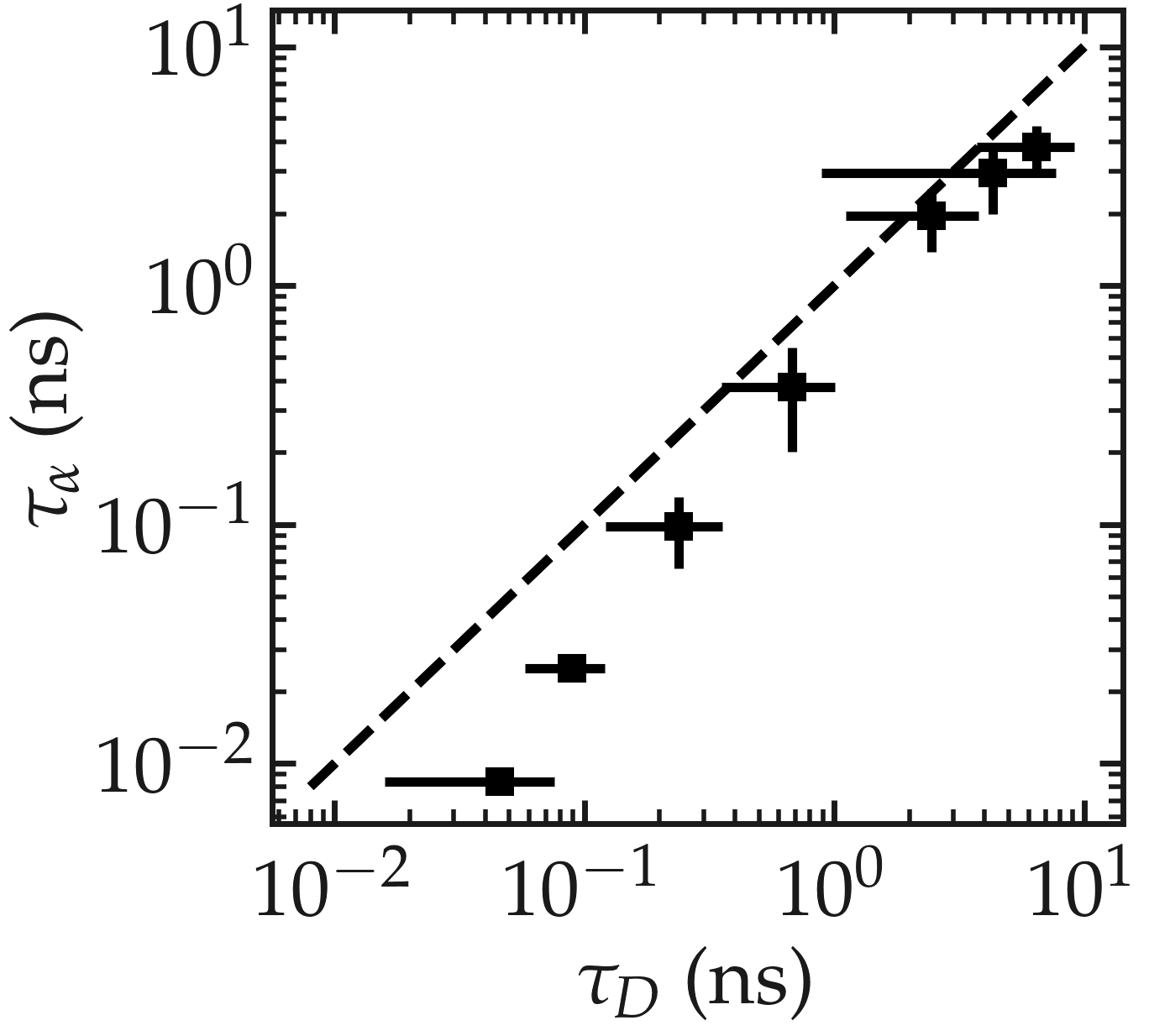}
\caption{Characteristic diffusion time $\tau_D = \sigma_\mathrm{Cs}^2/(6D)$ (red squares), where $\sigma_\mathrm{Cs} \approx 0.5$~nm is a representative ionic length scale (of the order of the cation diameter) and $D$ is the Cs$^+$ self-diffusion coefficient, as a function of the orientational relaxation time of the PEO chains, $\tau_\alpha$. The dashed line corresponds to $x = y$ (conditions: $C = 0.8$~M, $T = 60~^\circ\mathrm{C}$).}
\label{fig:timescomparison}
\end{figure}

\begin{figure}
\centering
\includegraphics[width=0.3\columnwidth]{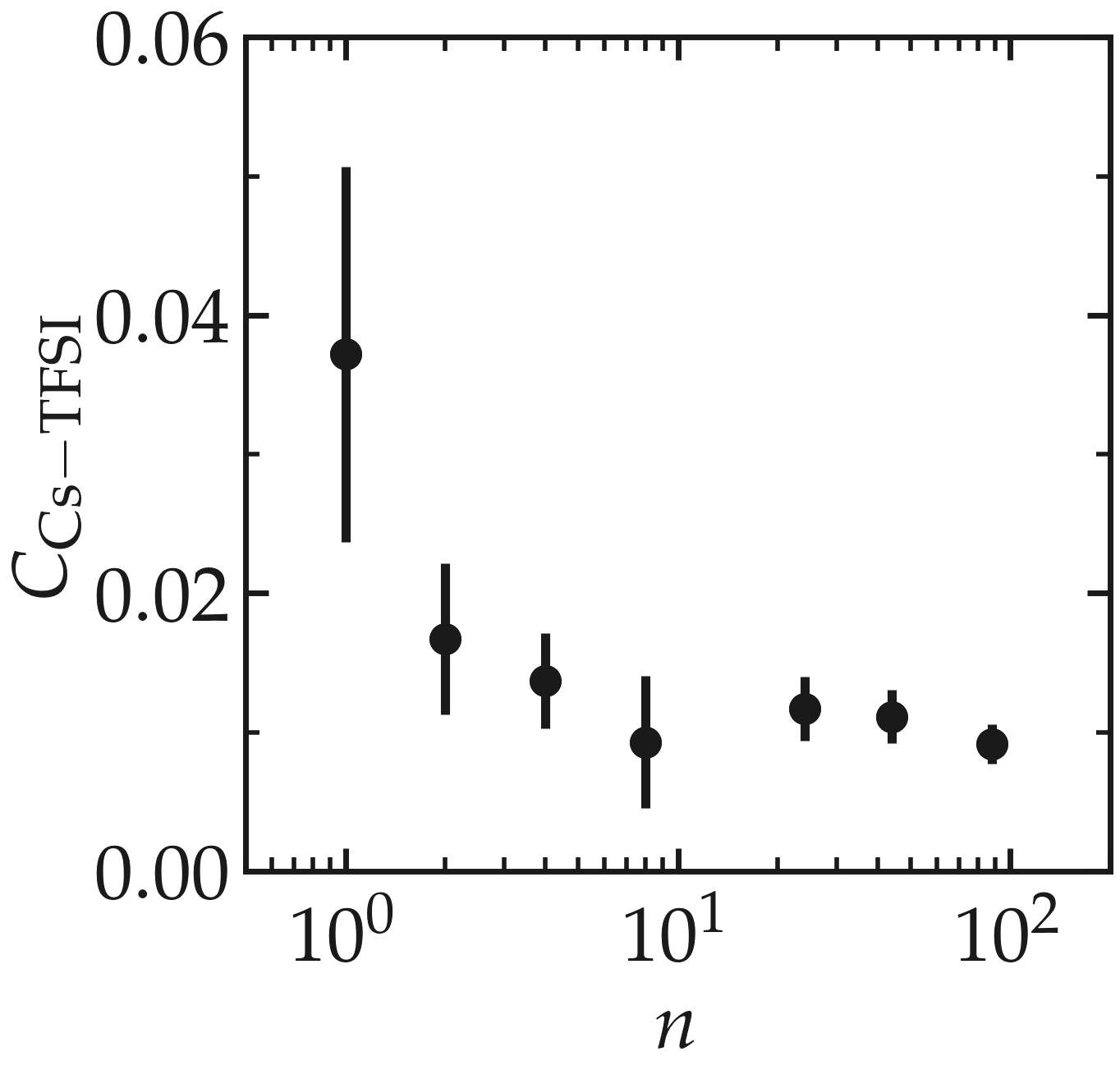}
\caption{Short-time displacement cross-correlation between Cs$^+$ and TFSI$^-$ [Limit $t \to 0$ of Eq.~\eqref{main-eq:cross-correlation} from the main text] 
as a function of $n$ (conditions: $C = 0.8$~M, $T = 60~^\circ\mathrm{C}$).}
\label{fig:cross-correlation}
\end{figure}

\begin{figure}
\centering
\includegraphics[width=0.3\columnwidth]{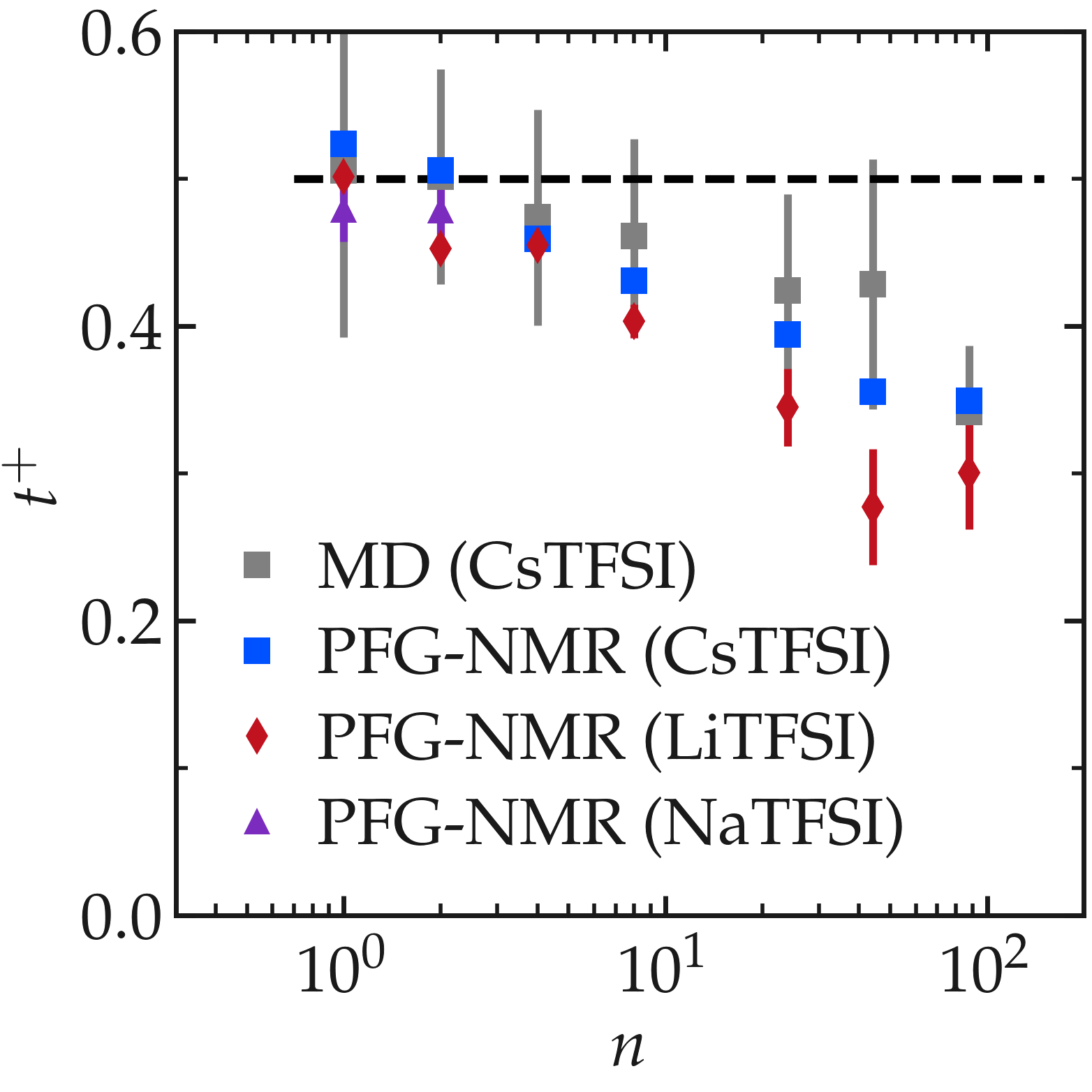}
\caption{Cation transference number, $t^+$, [Eq.~\eqref{main-eq:transference} from the main text] as a function of the degree of polymerization, $n$, from PFG-NMR and MD. The dashed line indicates $t^+ = 0.5$ (conditions: $C = 0.8$~M, $T = 60~^\circ\mathrm{C}$).}
\label{fig:transference}
\end{figure}

\bibliographystyle{ieeetr}
\bibliography{zotero-export}